\documentclass[APA,STIX1COL,doublespace]{WileyNJD-v2}

\articletype{research article}%

\received{26 April 2016}
\revised{6 June 2016}
\accepted{6 June 2016}

\usepackage{changes}
\usepackage{xcolor}

\begin{document}

\title{The X factor: A Robust and Powerful Approach to X-chromosome-Inclusive Whole-genome Association Studies}

\author[1]{Bo Chen}

\author[1]{Radu V. Craiu}

\author[1,2,3,4]{Lisa J. Strug}

\author[1,2]{Lei Sun*}

\authormark{Bo Chen \textsc{et al}}

\address[1]{\orgdiv{Department of Statistical Sciences}, \orgname{University of Toronto}, \orgaddress{\state{Ontario}, \country{Canada}}}

\address[2]{\orgdiv{Dalla Lana School of Public Health}, \orgname{University of Toronto}, \orgaddress{\state{Ontario}, \country{Canada}}}

\address[3]{\orgdiv{Department of Computer Science}, \orgname{University of Toronto}, \orgaddress{\state{Ontario}, \country{Canada}}}

\address[4]{\orgdiv{Program in Genetics and Genome Biology}, \orgname{The Hospital for Sick Children}, \orgaddress{\state{Ontario}, \country{Canada}}}

\corres{*Lei Sun, Department of Statistical Sciences, University of Toronto, Ontario, Canada. \email{sun@utstat.toronto.edu}}


\abstract[Summary]{The X-chromosome is often excluded from genome-wide association studies because of analytical challenges. Some of the problems, such as the random, skewed or no X-inactivation model uncertainty, have been investigated. Other considerations have received little to no attention, such as the value in considering non-additive and gene-sex interaction effects, and the inferential consequence of choosing different baseline alleles (i.e.\ the reference vs.\ the alternative allele). Here we propose a unified and flexible regression-based association test for X-chromosomal variants. We provide theoretical justifications for its robustness in the presence of various model uncertainties, as well as for its improved power when compared with the existing approaches under certain scenarios.  For completeness, we also revisit the autosomes and show that the proposed framework leads to a more robust approach than the standard method.  Finally, we provide supporting evidence by revisiting several published association studies. Supplementary materials for this article are available online.}


\maketitle

\section{Introduction}
\label{intro}

Genome-wide association studies (GWAS) are ubiquitous, delivering significant insights into the genetic determinants of complex traits over the past decade \citep{visscher17}. For this reason, it is surprising that it is not a common practice to include the X-chromosome in GWAS \citep{wise13, konig14}. The X-chromosome differs from the autosomes in that males have only one copy of the X-chromosome while females have two, and at any given genomic location one of the two copies in females may be silenced \citep{gendrel11}, referred to as X-chromosome inactivation (XCI).  The choice of the silenced copy could be random or skewed towards a specific copy \citep{wang14}.  These unique aspects lead to more complex analytic considerations for genetic association analysis of X-chromosomal variants, such as bi-allelic single nucleotide polymorphisms (SNPs). 

A bi-allelic SNP has two alleles, $r$ and $R$, of which one is the reference allele and the other is the alternative allele with allele frequency $f$. An autosomal SNP has three genotypes regardless of sex, namely $G = (rr, rR, RR)$.  In association analysis of an autosomal SNP, the common practice is to simply model a binary or continuous phenotype $Y$ as an additive function of the number of copies of the non-baseline allele present in $G$; that is, coding $G$ additively as $G_A=(0,1, 2)$. Here, without loss of generality, $r$ is chosen to be the baseline allele in a statistical model and $R$ the non-baseline allele. When $Y$ is binary, this regression-based additive test is also equivalent to the Cochran-Armitage trend test \citep{wellek12}. Although both dominant and recessive genetic models of inheritance are possible, among these one degrees of freedom (1 d.f.) models, a common practice for GWAS is to use the additive model, because it has reasonable power to detect both additive and dominance effects at a causal variant, and at variants in linkage disequilibrium (LD) with the causal variant \citep{hill08, bush12}.  An alternative parameterization is the 2 d.f.\ genotypic model that includes both the additive $G_A=(0,1, 2)$ term and the dominance $G_D=(0,1,0)$ term. In the case of recessive genetic inheritance, \cite{zhou17} showed that the 2 d.f.\ genotypic test outperforms the 1 d.f.\ additive test for binary outcomes, and \cite{dizier17} reached the same conclusion for continuous traits. In the case of additive genetic inheritance being true, the genotypic test is known to be less powerful than the additive test due to the increased d.f., which is unnecessary. The preferred test for unknown genetic inheritance in terms of power and robustness to different genetic models is, however, not clear across different true genetic effect sizes, sample sizes and significance levels.

For an X-chromosomal SNP, the most commonly used approach assumes additivity and X-chromosome inactivation (XCI). However, recent work \citep{tukiainen17} showed that up to one third of X-chromosomal genes are expressed from both the active and inactive X-chromosomes in female cells, with varying degrees of `escape' from inactivation between genes and individuals. Several additional points also require attention. Table \ref{tab1} describes eight analytical considerations and challenges (C1-C8) present in an X-chromosome-inclusive GWAS, including a method's suitability for analyzing both binary and continuous traits (C1), which is related to the type of method used, i.e.\ genotype-based or allelic association tests (C2); the (under-appreciated) consequence of the choice of the baseline allele on association analysis of an X-chromosomal SNP (C3); the importance of including sex as a covariate (C4) and its analytical connection with C3; the value in considering gene-sex interaction effect (C5) and its connection with the assumption of XCI (C6); and the assumption of random vs.\ skewed XCI (C7) and its connection with non-additive effects (C8).

\begin{table}
\begin{center}
\caption{\bf Eight analytical considerations and challenges, C1-C8, present in X-chromosome-inclusive association studies.}
\label{tab1}
\resizebox{\textwidth}{!}{
\begin{tabular}{lll} \hline \hline

\multicolumn{1}{m{8.5cm}}{\bf Problem} 
&\multicolumn{1}{m{6.5cm}} {\bf Solution} &\multicolumn{1}{m{2cm}} {\bf Relevant Sections}\\ \hline

\multicolumn{3}{l}{\bf C1: Quantitative traits vs.\ binary outcomes}\\
\multicolumn{3}{l}{\bf C2: Genotype-based vs.\ allele-based association methods}\\ 
\multicolumn{1}{m{8.5cm}}{Allele-based association tests, comparing allele frequency differences between cases and controls, are locally most powerful. However, they analyze binary outcomes only and are sensitive to the Hardy-Weinberg equilibrium (HWE) assumption \citep{sasieni97}.}
&\multicolumn{1}{m{6.5cm}} {Genotype-based regression models, $Y$-on-$G$, support various types of outcome data, account for covariate effects with ease, and are robust to the HWE assumption.} 
&\multicolumn{1}{m{2cm}} {Sections \ref{intro} and \ref{Xchr}}\\ \hline

\multicolumn{3}{l}{\bf C3: The choice of the baseline allele for association analysis, $r$ vs.\ $R$} \\
\multicolumn{1}{m{8.5cm}}{For the autosomes, switching the two alleles does not affect the association inference. Is this true for the X-chromosome?}
&\multicolumn{1}{m{6.5cm}}{ It is {\it not} always true for the X-chromosome, unless $S$ is included in the model.}
&\multicolumn{1}{m{2cm}} {Sections \ref{s:coding} and \ref{includingS}, and C4 below} \\  \hline

\multicolumn{3}{l}{\bf C4: Sex as a covariate vs.\ no $S$ main effect}\\ 
\multicolumn{1}{m{8.5cm}}{Unlike the autosomes, sex is a confounder when analyzing the X-chromosome for traits exhibiting sexual dimorphism (e.g.\ height and weight). Even for the autosomes, sex can be a confounder if allele frequencies differ significantly between males and females.} 
&\multicolumn{1}{m{6.5cm}}{To maintain the correct type I error rate control, the sex main effect must be considered particular when analyzing the X-chromosome. The resulting association test is also invariant to the choice of the baseline allele.}
&\multicolumn{1}{m{2cm}} {Section \ref{includingS} and C3 above}\\  \hline

\multicolumn{3}{l}{\bf C5: Gene-sex interaction vs.\ no $G\times S$ interaction effect} \\ 
\multicolumn{1}{m{8.5cm}}{Gene-sex interaction might exist, but there is a concern over loss of power due to increased degrees of freedom. In addition, what is the interpretation of gene-sex interaction effect in the presence of X-inactivation?} 
&\multicolumn{1}{m{6.5cm}}{Under no interaction, power loss of modelling interaction is capped at 11.4\%. Models including the $G\times S$ covariate also lead to tests invariant to the assumption of X-chromosome inactivation status.}
&\multicolumn{1}{m{2cm}} {Sections \ref{includingGS} and \ref{section:powerX}, and C6 below} \\ \hline

\multicolumn{3}{l}{\bf C6: X-chromosome inactivation (XCI) vs.\ no XCI}\\
\multicolumn{1}{m{8.5cm}}{XCI occurs if one of the two alleles in a genotype of a female is silenced. Individual-level XCI status requires additional biological information that are not typically available to genetic association studies.  Assuming XCI or no XCI at the sample level leads to different genotype coding strategies (Table \ref{coding}), and it was thought that this will always lead to different association results.}
&\multicolumn{1}{m{6.5cm}}{XCI uncertainty implies sex-stratified genetic effect which can be analytically represented by the $G\times S$ interaction effect. Teasing apart these different biological phenomenons require other `omic' data and additional analyses.} 
&\multicolumn{1}{m{2cm}} {Sections \ref{includingGS} and \ref{discussion}, and C5 above} \\ \hline

\multicolumn{3}{l}{\bf C7: If XCI, random vs.\ skewed X-inactivation}\\
\multicolumn{1}{m{8.5cm}}{If the choice of the silenced allele in females is skewed towards a specific allele, the average effect of the $rR$ genotype is no longer the average of those of $r$ and $R$.}
&\multicolumn{1}{m{6.5cm}}{XCI skewness is statistically equivalent to a dominance genetic effect.}
&\multicolumn{1}{m{2cm}} {Section \ref{includingD}, and C8 below} \\ \hline

\multicolumn{3}{l}{\bf C8: {Dominance effect vs. no $G_D$ dominance effect}}\\
\multicolumn{1}{m{8.5cm}}{For both the autosomes and X-chromosome, the most common practice is to use the additive test which has better power than the genotypic test under (approximate) additivity, but it cannot capture dominance effects. The exact trade-off, however, is not clear.}
&\multicolumn{1}{m{6.5cm}}{We provide analytical and empirical evidences supporting the use of genotypic model when analyzing either the autosomes or X-chromosome. For an X-chromosomal variant, including the dominance effect term has the added benefit of resolving of the skewed X-inactivation uncertainty issue.}
&\multicolumn{1}{m{2cm}} {Sections \ref{includingD}, \ref{section:powerX} and \ref{application}, and C7 above} \\ \hline\hline

\end{tabular}
}
\end{center}
\end{table}

Several association methods have been developed for the X-chromosome, and they are computationally efficient for conducting X-chromosome-wide association analysis. However, each method solves only some of the C1-C8 challenges.  For example, \cite{zheng07} considered only binary outcomes for which both genotype- and allele-based association tests are applicable. The classical allelic association test, comparing allele frequencies between case and control groups, is locally most powerful but sensitive to the Hardy-Weinberg equilibrium (HWE) assumption and not applicable to continuous traits \citep{sasieni97, zheng08,zhang21}. \cite{clayton08,clayton09} discussed analytical strategies assuming the X-chromosome is always inactivated. \cite{hickey11} and \cite{loley11} performed simulation studies, each providing a thorough method comparison, e.g. between tests of \cite{zheng07} and \cite{clayton08}. \cite{konig14} gave detailed guidelines for including the X-chromosome in GWAS, recommending different tests for different model assumptions (e.g.\ presence or absence of an interaction effect or XCI), but it is difficult to validate these assumptions in practice. \cite{gao15} developed a toolset for conducting X-chromosome association studies, implementing some of the existing methods. More recently \cite{chen17} improved sex-stratified analysis by eliminating genetic model assumptions, but their method is limited to analyzing genetic main effects on binary traits. Focusing on XCI uncertainty, \cite{wang14} proposed a frequentist maximum likelihood solution to deal with no, random or skewed X-inactivation, and in their follow-up work \cite{wang17} provided a model selection method. In contrast, \cite{chen20} applied the Bayesian model averaging principle \citep{draper} to deal with the XCI uncertainty problem. However, these approaches assumed additive genetic effects. The value in considering dominance and gene-sex interaction effects, and the inferential consequence of defining different baseline allele (i.e.\ the reference or the alternative allele) when analyzing an X-chromosomal SNP, have received little to no attention. 
 
Here we propose a theoretically justified and robust X-chromosome association method that can simultaneously deal with all eight challenges (C1--C8) outlined in Table \ref{tab1}. We emphasize the robustness of the proposed method to genetic assumptions as our understanding is evolving. For example, although most published X-chromosome-inclusive GWAS assumed XCI, recent work has shown that up to a third of genes `escape' XCI \citep{tukiainen17}. 

The proposed method is regression- and genotype-based (robust to departure from HWE), analyzing either a continuous or binary trait while adjusting for covariate effects. The recommended test has three degrees of freedom, including both additive and dominance genetic effects, as well as a gene-sex interaction effect. We show analytically why the proposed method is robust to the various model uncertainties, including no, random or skewed X-chromosome inactivation, as well as the choice of the baseline allele. Desirably, the power of the proposed test is robust to different alternative genetic models, despite its increased degrees of freedom over a simple additive test. We note that the work here focuses on efficient association testing, not parameter estimation or model selection which requires additional biological data \citep{busque96}.

We first present our main theory to address the eight challenges associated with X-chromosome-inclusive GWAS in Section \ref{Xchr}. We then provide analytical results of power study across all possible genetic models, sample sizes and type I error rates, as well as empirical results from simulation studies in Section \ref{section:powerX}. For methodology completeness, this section also briefly discusses merit of the genotypic model in the familiar context of analyzing autosomal SNPs. We then provide corroborating evidence from several applications in favour of the proposed approach in Section \ref{application}. Finally we discuss the limitations of our approach and possible future work in Section \ref{discussion}. 

\section{Method for X-chromosome-inclusive association analysis}
\label{Xchr}

The proposed method relies on the generalized linear model \citep{mccullagh89} as it is flexible, analyzing both binary and continuous traits (C1 of Table \ref{tab1}). As a result, the method is a genotype-based approach (C2) that is robust to the assumption of Hardy-Weinberg equilibrium by regressing the phenotype data ($Y$) on genetic data ($G$) while accounting for other covariate effects.

For robust and powerful association analysis of a bi-allelic X-chromosomal SNP, we recommend the following model,
\begin{equation}\label{eq1}
g(E(Y))=\beta_0+\beta_S S+\beta_A G_A+\beta_{D} G_D + \beta_{GS} GS,
\end{equation}
and the corresponding 3 d.f.\ test, jointly testing
\begin{equation}\label{eq2}
H_0: \beta_A=\beta_D=\beta_{GS}=0,
\end{equation}
where notations for the covariates are defined in Table \ref{coding}. Other relevant covariates such as environmental factors ($E$'s) should also be included in the model but omitted here for notation simplicity.

We show later (a) why the association result from the proposed approach is invariant to the different $G_A$ (e.g.\ $G_{A,R,I}$ or $G_{A,r,N}$) and $GS$ (e.g. $GS_{R}$ or $GS_{r}$) coding schemes as defined in Table \ref{coding}, and (b) why the proposed method also solves the C3-C8 issues simultaneously. But before we do so, we first provide more details about the notations presented in Table \ref{coding}.

\subsection{X-chromosome specific genotype and covariate coding schemes}
\label{s:coding}

Table \ref{coding} summarizes the various covariate coding schemes for analyzing an X-chromosomal SNP, when considering all the analytical challenges outlined in Table \ref{tab1}. Note that when the choice of the baseline allele is varied (i.e.\ either $r$ or $R$) and the XCI status is unknown, there are four ways to code the additive covariate $G_A$, and two ways to code the gene-sex interaction covariate $GS$. The specific coding for sex does not have an impact on our proposed method. In Table \ref{coding}, a female is coded as 0 and a male as 1, and the interaction $G_D \times S$ term vanishes. If a female were coded as 1 and a male as 0, then $G_D \times S$ is the same as $G_D$. Thus, in either case it is redundant to include $G_D \times S$ in our proposed regression model.

\begin{table}
\begin{center}
\caption{{\bf Covariate coding schemes for examining the additive, dominance, gene-sex interaction, and sex effects under different assumptions of the X-chromosome inactivation status and the choice of the baseline allele}. The subscripts $A$ and $D$ represent additive and dominance effects, $R$ or $r$ represents the non-baseline allele of which we count the number of copies present in a genotype, and $I$ or $N$ denotes X-chromosome inactivated or not inactivated.}
\bigskip
\label{coding}
\begin{tabular}{c|c|c|c|c|c|c|c|c} \hline \hline
& & & X-chromosome & \multicolumn{5}{c}{Coding Schemes} \\ 
Effect & Covariate & Non-Baseline  &  Inactivation &\multicolumn{3}{c}{(Females)}& \multicolumn{2}{c}{(Males)} \\ \cline{5-7} \cline{8-9}
Interpretation& Notation & Allele & (XCI) Status & $rr$ & $rR$ & $RR$ & $r$ & $R$ \\  \hline
& $G_{A,R,I}$ & $R$ & Yes & 0 & 0.5 & 1 & 0 & 1 \\ \cline{2-9}
Additive & $G_{A,r,I}$ & $r$ & Yes & 1 & 0.5 & 0 & 1 & 0 \\ \cline{2-9}
$G_A$ & $G_{A,R,N}$ & $R$ & No & 0 & 1 & 2 & 0 & 1 \\ \cline{2-9}
& $G_{A,r,N}$ & $r$ & No & 2 & 1 & 0 & 1 & 0 \\ \hline 
Dominance $G_D$ & $G_D$ & Either & Either & 0 & 1 & 0 & 0 & 0 \\ \hline
Gene-Sex Interaction & $GS_{R}$ & $R$ & Either & 0 & 0 & 0 & 0 & 1 \\ \cline{2-9}
$GS=G_A \times S$ & $GS_{r}$ & $r$ & Either & 0 & 0 & 0 & 1 & 0 \\ \hline \hline
Sex $S$ & $S$ & Either & Either & 0 & 0 & 0 & 1 & 1 \\ \hline \hline 
\end{tabular}
\end{center}
\end{table}

Using the notations in Table \ref{coding}, it is immediately clear why the choice of the baseline allele (C3) matters for association analysis of an X-chromosomal SNP. Under no XCI, if $r$ were assumed to be the baseline allele there would be one copy of allele $R$ in genotype $rR$ of a female, and  $R$ of a male. Thus, genotypes $rR$ and $R$ would be grouped together for association analysis. However, if $R$ were chosen to be the baseline allele, genotypes $rR$ and $r$ would be grouped together, resulting in different inference. In contrast, the choice of the baseline allele does not affect association evidence when analyzing an autosomal SNP. It is well-known that although the estimate of the effect size changes direction, the magnitude of the association remains the same when analyzing an autosomal SNP. But, this is not always true when analyzing an X-chromosomal SNP.

\subsection{Sex as a confounder (C4) and its connection with the choice of the baseline allele (C3)}
\label{includingS}

Sex is a confounder for phenotype-genotype association analysis of an X-chromosomal SNP for traits displaying sexual dimorphism.  When sex, but not the SNP, is associated with a trait of interest, omitting sex in the analysis leads to false positives. This is because sex is inherently associated with the genotypes of an X-chromosomal SNP (Table \ref{coding}); see \cite{ozbek18} for empirical evidence from simulation studies.  Thus, accuracy of a test provides the first argument for always including $S$ as a covariate in association analysis of an X-chromosomal SNP.

The second advantage of modelling the $S$ main effect is more subtle.  As shown in Table \ref{coding}, the coding of $G_A$ depends on the choice of the baseline allele (i.e.\ $R$ or $r$) and the X-inactivation status ($I$ for XCI and $N$ for no XCI), resulting in a total of four different ways of coding the five genotype groups, namely $G_{A,R,I}=(0,0.5,1,0,1)'$, $G_{A,r,I}=(1,0.5,0,1,0)'$, $G_{A,R,N}=(0,1,2,0,1)'$, and $G_{A,r,N}=(2,1,0,1,0)'$. Furthermore, $G_{A, R, N}$ and $G_{A, r, N}$ yield different test statistics, because the two coding schemes lead to different groupings of the genotypes as discussed in \ref{s:coding}. Note that, in contrast to $G_{A,R,I}=1-G_{A,r,I}$ under XCI, under no XCI there is no linear transformation that makes $G_{A, R, N}$ and $G_{A, r, N}$ equivalent. An inference that is invariant to the coding choices may seem difficult, but we show that this is achievable for models that include sex as a covariate.

\begin{theorem}
\label{transformation}
Let $\mathcal{M}_1$ and $\mathcal{M}_2$ be two generalized linear models \citep{mccullagh89} with the same link function $g$, $g(E(Y))=X_1\boldsymbol{\beta_1}$ and $g(E(Y))=X_2\boldsymbol{\beta_2}$, where $Y$ is the response vector of length $n$, $X_1$ and $X_2$ are two $n \times p$ design matrices, and $\boldsymbol{\beta_1}$ and $\boldsymbol{\beta_2}$ are the corresponding parameter vectors of length $p$. Let $X_1=(X_{11}, X_{12})$, where $X_{11}$ and $X_{12}$ are $n \times (p-q)$ and $n \times q$ matrices corresponding to, respectively, the $(p-q)$ secondary covariates not being tested and the $q$ primary covariates of interest, and similarly for $X_2=(X_{21}, X_{22})$, and partition the regression coefficients accordingly as $\boldsymbol{\beta_1}=(\beta_{11}', \beta_{12}')'$ and $\boldsymbol{\beta_2}=(\beta_{21}', \beta_{22}')'$. If there exists an invertible $p\times p$ matrix 
$$T=\left(\begin{array}{cc}
                   T_{11} & T_{12} \\
                   0 & T_{22} \\
           \end{array}\right), \mbox{ such that} \:\:\: X_2=X_1\: T, $$
where $T_{11}$ and $T_{22}$ are, respectively, invertible $(p-q) \times (p-q)$ and $q \times q$ matrices, then any of the Wald, Score or LRT tests for testing 
$$H_0: \beta_{12}=0 \:\: \text{and} \:\:  H_0: \beta_{22}=0$$ 
are identical under the two models $\mathcal{M}_1$ and $\mathcal{M}_2$, resulting in the same association inference for evaluating the $q$ primary covariates of interest. Note that given the structure of matrix $T$, $X_2=X_1\: T$ implies $X_{21} = X_{11}\: T_{11}$.
\end{theorem}

We provide the proof of Theorem \ref{transformation} in Web Appendix A. Here we emphasize that the two sets of $q$ primary covariates being tested, $X_{22}$ and $X_{12}$, are {\it not} required to be linear transformation of each other, e.g., between $G_{A,R,N}=(0,1,2,0,1)'$ and $G_{A,r,N}=(2,1,0,1,0)'$. Instead, $X_{21}$ and $X_{11}$, corresponding to the $p-q$ secondary covariates (including the unit vector if modelling the intercept), that are {\it not} being tested must be invertible linear transformations of each other, $X_{21} = X_{11}\: T_{11}$, in addition to $X_2=X_1 \: T$. This result may seem surprising, but the two requirements imply that the two design matrices are equivalent to each other either in general or under the null, resulting in identical F-test statistics; see Web Appendix A for technical details.    

In our setting when sex is included in the model, consider only the additive effect for the moment, $g(E(Y))=\beta_0+\beta_SS+\beta_AG_A$. Then the two design matrices, corresponding to $r$ or $R$ being the baseline allele and under no XCI, are 
$$X_1=\left(\begin{array}{ccc}
                   1 & 0 & 0 \\
                   1 & 0 & 1 \\
                   1 & 0 & 2 \\
                   1 & 1 & 0 \\
                   1 & 1 & 1 \\     
           \end{array}\right) \: \mbox{ and} \:\:  
X_2=\left(\begin{array}{ccc}
                   1 & 0 & 2 \\
                   1 & 0 & 1 \\
                   1 & 0 & 0 \\
                   1 & 1 & 1 \\
                   1 & 1 & 0 \\     
           \end{array}\right).$$
In this case, $T_{11} = \left(\begin{array}{cc}
                   1 & 0 \\
                   0 & 1 \\
\end{array}\right)$, $T_{12}= (2, -1)'$ and $T_{22}={-1}$ satisfy the two requirements. Thus, even though $G_{A,R,N}=(0,1,2,0,1)'$ and $G_{A,r,N}=(2,1,0,1,0)'$ are not linked by a linear transformation, Theorem \ref{transformation} allows us to conclude that a Wald, Score or LRT test of $H_0: \beta_A=0$ is invariant to the two $G_A$ coding schemes $G_{A,R,N}$ and $G_{A,r,N}$, if sex is included as a covariate.

Note that the known result that two tests are equivalent to each other if $X_{12}$ and $X_{22}$, corresponding to $q$ primary covariates, are linear transformation of each other is a special case of Theorem \ref{transformation}, where all elements except the first row of $T_{12}$ are zero; the exception allows for a location shift. For example, under the XCI assumption, $X_{12}=G_{A,R,I}=(0,0.5,1,0,1)'$ and $X_{22}=G_{A,r,I}=(1,0.5,0,1,0)'$, and $X_{22}=1-X_{12}$. Thus, $T_{11} = \left(\begin{array}{cc}
                   1 & 0 \\
                   0 & 1 \\
\end{array}\right)$, $T_{12}= (1, 0)'$ and $T_{22}={-1}$ satisfy the requirements.

At this point in the methodology development, the preferred model $g(E(Y))=\beta_0+\beta_SS+\beta_AG_A$ controls the type I error rate and is invariant to the choice of the baseline allele.  However, in practice the XCI status is unknown and if we assume there is XCI, $G_{A,R,I}=(0,0.5,1,0,1)'$ and $G_{A,r,N}=(2,1,0,1,0)'$, and
$$X_1=\left(\begin{array}{ccc}
                   1 & 0 & 0 \\
                   1 & 0 & 0.5 \\
                   1 & 0 & 1 \\
                   1 & 1 & 0 \\
                   1 & 1 & 1 \\     
           \end{array}\right) \:\: \mbox{and} \:\:   
X_2=\left(\begin{array}{ccc}
                   1 & 0 & 2 \\
                   1 & 0 & 1 \\
                   1 & 0 & 0 \\
                   1 & 1 & 1 \\
                   1 & 1 & 0 \\     
           \end{array}\right).$$
In this case, it is not difficult to show that a matrix $T$ satisfying the requirements of Theorem \ref{transformation} does not exist, {because $Rank(X_1)<Rank((X_1,X_2))$ implies that the linear system $X_2=X_1 \: T$ has no solution,} and the XCI uncertainty remains a challenge.

\subsection{Gene-sex interaction effect (C5) and its connection with unknown X-chromosome inactivation status (C6)}
\label{includingGS}

Throughout the paper, we define the $GS$ interaction term as $G_A \times S$.  Depending on the choice of the baseline allele, $GS$ has two different codings, namely $GS_{R}$ and $GS_{r}$ (Table \ref{coding}).  In the previous section, we have shown that when $S$ is included in the model, i.e. $g(E(Y))=\beta_0+\beta_SS+\beta_AG_A$, the choice of the baseline allele is no longer of a concern if we test $H_0: \beta_A=0$ within a particular XCI assumption.  Interestingly, when both $S$ and $GS$ are included in the model, $g(E(Y))=\beta_0+\beta_SS+\beta_AG_A+\beta_{GS}GS$, by applying Theorem \ref{transformation} again, testing $H_0: \beta_A=\beta_{GS}=0$ is statistically equivalent between the different choices of the baseline allele {\it and} the assumption of the XCI status. For example, consider 
$$X_1=\left(\begin{array}{cccc}
                   1 & 0 & 0    & 0\\
                   1 & 0 & 0.5 & 0\\
                   1 & 0 & 1    & 0\\
                   1 & 1 & 0    & 0\\
                   1 & 1 & 1    & 1\\     
           \end{array}\right) \:\: \mbox{and} \:\: 
X_2=\left(\begin{array}{cccc}
                   1 & 0 & 2 & 0 \\
                   1 & 0 & 1 & 0 \\
                   1 & 0 & 0 & 0\\
                   1 & 1 & 1 & 1\\
                   1 & 1 & 0 & 0\\     
           \end{array}\right),$$
respectively, for a model assuming XCI and choosing $r$ as the baseline allele (i.e.\ tracking the number of copies of allele $R$), and for a model assuming no XCI and choosing $R$ as the baseline allele, we can show that 
$$T_{11} = \left(\begin{array}{cc}
                   1 & 0 \\
                   0 & 1 \\
\end{array}\right),\:\: 
T_{12} = \left(\begin{array}{cc}
                   2 & 0 \\
                   -1 & 1 \\
\end{array}\right) \:\: \mbox{and} \:\:  
T_{22} = \left(\begin{array}{cc}
                   -2 & 0 \\
                   1 & -1 \\
\end{array}\right)$$ 
satisfy the linear transformation requirements of Theorem \ref{transformation}.  That is, for association analysis of an X-chromosomal SNP, testing $H_0: \beta_A=\beta_{GS}=0$ based on $g(E(Y))=\beta_0+\beta_SS+\beta_AG_A +\beta_{GS} GS$ is invariant to the choice of the baseline allele and the assumption of the X-inactivation status. Figure \ref{tran} summarizes the equivalency between the design matrices that correspond to the different coding schemes studied so far; all the theoretical results have been confirmed empirically via simulations. 

\begin{figure}
\vspace{2mm}
\begin{center}\setlength{\unitlength}{1in}
\begin{picture}(0,0)
\put(-3.1,-0.1){(a)}
\put(-2.8,0){$g(E(Y))= \beta_0+\beta_AG_A$}
\put(-2.8,-0.2){$H_0: \beta_A=0$}
\put(-3.1,-0.6){(b)}
\put(-2.8,-0.5){$g(E(Y))= \beta_0+\beta_AG_A+\beta_DG_D$}
\put(-2.8,-0.7){$H_0: \beta_A=\beta_D=0$}
\put(0.1,-0.1){(a)}
\put(0.4,0){$g(E(Y))=\beta_0+\beta_SS+\beta_AG_A$}
\put(0.4,-0.2){$H_0: \beta_A=0$}
\put(0.1,-0.6){(b)}
\put(0.4,-0.5){$g(E(Y))= \beta_0+\beta_SS+\beta_AG_A+\beta_DG_D$}
\put(0.4,-0.7){$H_0: \beta_A=\beta_D=0$}
\put(-1.5,-2.1){(a)}
\put(-1.2,-2.0){$g(E(Y))=\beta_0+\beta_SS+\beta_AG_A+\beta_{GS}GS$}
\put(-1.2,-2.2){$H_0: \beta_A=\beta_{GS}=0$}
\put(-1.5,-2.6){(b)}
\put(-1.2,-2.5){$g(E(Y))=\beta_0+\beta_SS+\beta_AG_A+\beta_DG_D+\beta_{GS}GS$}
\put(-1.2,-2.7){$H_0: \beta_A=\beta_D=\beta_{GS}=0$}
\put(-2.8,-1.1){\framebox(0.6,0.2){$G_{A,R,I}$}}
\put(-1.3,-1.1){\framebox(0.6,0.2){$G_{A,r,I}$}}
\put(-2.8,-1.6){\framebox(0.6,0.2){$G_{A,R,N}$}}
\put(-1.3,-1.6){\framebox(0.6,0.2){$G_{A,r,N}$}}
\put(0.6,-1.1){\framebox(0.8,0.2){$S,G_{A,R,I}$}}
\put(2.1,-1.1){\framebox(0.8,0.2){$S,G_{A,r,I}$}}
\put(0.6,-1.6){\framebox(0.8,0.2){$S,G_{A,R,N}$}}
\put(2.1,-1.6){\framebox(0.8,0.2){$S,G_{A,r,N}$}}
\put(-1.2,-3.1){\framebox(1.2,0.2){$S,G_{A,R,I},GS_{R}$}}
\put(0.4,-3.1){\framebox(1.2,0.2){$S,G_{A,r,I},GS_{r}$}}
\put(-1.2,-3.6){\framebox(1.2,0.2){$S,G_{A,R,N},GS_{R}$}}
\put(0.4,-3.6){\framebox(1.2,0.2){$S,G_{A,r,N},GS_{r}$}}
\put(-2.2,-1.0){\line(1,0){0.9}}
\put(1.4,-1.0){\line(1,0){0.7}}
\put(1.4,-1.5){\line(1,0){0.7}}
\put(0,-3.0){\line(1,0){0.4}}
\put(0,-3.5){\line(1,0){0.4}}
\put(-0.55,-3.1){\line(0,-1){0.3}}
\put(0.95,-3.1){\line(0,-1){0.3}}
\put(0,-3.1){\line(4,-3){0.4}}
\put(0,-3.4){\line(4,3){0.4}}
\end{picture}
\end{center}
\vspace{3.6in}
\caption{{\bf Equivalency between different regression models for association analysis of an X-chromosomal bi-allelic SNP.} The subscript $R$ or $r$ represents the non-baseline allele of which we count the number of copies present in a genotype, and $I$ or $N$ denotes X-chromosome inactivated or not inactivated; see Table \ref{coding} for additional covariate coding details. Two groups of coding connected by a line if there is an invertible linear transformation between the design matrices as specified in Theorem \ref{transformation}, and the resulting test statistics for testing the specified $H_0$ will be identical to each other. Part (a) corresponds to models and tests without the dominance $G_D$ covariate, and part (b) corresponds to models and tests with $G_D$ included.  Inclusion of $G_D$ has no effect on the linear relationships established in part (a), because coding of $G_D$ in Table \ref{coding} is invariant to the choice of the baseline allele or the XCI status. However, $G_D$ effect is statistically equivalent to skewed XCI as shown in section \ref{includingD}.}
\label{tran}
\end{figure}

\subsection{Random vs.\ skewed X-inactivation (C7) and its connection with genetic dominance effect (C8)}
\label{includingD}

Similar to analyzing an autosomal SNP, the first reason for modelling the dominance effect is to capture potential departure from additivity; see Section \ref{section:powerX} for additional discussion. For an X-chromosomal SNP, another important reason is that the dominance effect can also  {\it statistically} capture skewness of X-inactivation, if present. 

Intuitively, if we assume the effects of $rr$ and $RR$ to be, respectively, 0 and 1, the effect of $rR$ will be either 0 or 1 for each individual, depending on the inactivated allele of the sample collected. If the two alleles are equally likely to be inactivated (i.e.\ random XCI) across all individuals, the average statistical effect of $rR$ is 1/2. If $r$ is more (or less) likely to be inactivated (i.e.\ skewed XCI), the average effect of $rR$ is greater (or less) than 1/2. However, this XCI skewness is {\it analytically} equivalent to a dominance effect (i.e.\ effect of $rR$ deviating from 1/2), even though dominance effect is at the population level whereas skewed XCI is a sample-specific property. This analytical equivalency also shows that knowing the true underlying biological model requires more than the standard GWAS data.    

Table \ref{summary} summarizes the statistical behaviours of all the regression models and corresponding tests discussed in this section. Notably, jointly testing $H_0: \beta_A=\beta_D=\beta_{GS}=0$, based on the $g(E(Y))=\beta_0+\beta_S S+\beta_A G_A+\beta_{D} G_D + \beta_{GS} GS$ model $M_4$, ensures that the inference is invariant to the assumptions of the XCI status and baseline allele, and accounts for dominance effect and XCI-skewness if present.

\begin{table}
\begin{center}
\caption{{\bf Properties of different regression models in the presence of the eight analytical challenges, as detailed in Table \ref{tab1}.} Whole-genome considerations such as C1 (continuous vs. binary traits) and C2 (Hardy-Weinberg equilibrium vs.\ disequilibrium) are naturally dealt with by the genotype-based regression approach. X-chromosome-specific considerations include C3 (choice of the baseline allele), C4 (sex as a confounder and type I error control), C5 (gene-sex interaction), C6 (X-chromosome inactivation (XCI) vs. no XCI), C7 (random vs. skewed XCI), and C8 (the dominance effect). In the table, $\times$ indicates a problem for the corresponding model and test, and $\surd$ means no problem. Relevant covariates $E$'s should be included in the model but omitted here for notation simplicity. Joint testing of $H_0: \beta_A=\beta_D=\beta_{GS}=0$ based on $M_4$ is the recommended, most robust approach; see Figure \ref{powerX} and Web Figure S5 for power comparisons among $M_1$--$M_4$.}
\bigskip
\label{summary}
\begin{tabular}{l|l|c|c|c} \hline \hline
Model, $g(E(Y))=$ & Testing $H_0:$ & C3/C4 & C5/C6 & C7/C8 \\ \hline
$M_0: \beta_0+\beta_AG_A$ & $\beta_A=0$ & $\times$ & $\times$ & $\times$ \\ \hline 
$M_1: \beta_0+\beta_S S+\beta_A G_A$ & $\beta_A=0$ & $\surd$ & $\times$ & $\times$ \\ \hline 
$M_2: \beta_0+\beta_S S+\beta_A G_A+\beta_{D} G_D$ & $\beta_A=\beta_D=0$ & $\surd$ & $\times$ & $\surd$ \\ \hline 
$M_3: \beta_0+\beta_S S+\beta_A G_A+\beta_{GS} GS$ & $\beta_A=\beta_{GS}=0$ & $\surd$ & $\surd$ & $\times$ \\ \hline 
$M_4: \beta_0+\beta_S S+\beta_A G_A+\beta_{D} G_D + \beta_{GS} GS$  & $\beta_A=\beta_D=\beta_{GS}=0$ & $\surd$ & $\surd$ & $\surd$ \\ \hline \hline
\end{tabular}
\end{center}
\end{table}
 
\section{Analytical and Simulation-based Method Evaluation}
\label{section:powerX}

The proposed method is easy-to-implement and has good type I error control, because regression-based approach is known to be well behaved, as long as sample size is not too small and allele frequency is not too low, which are satisfied by most genome-wide association studies of common variants. Thus, we focus on evaluating power of the proposed method.  We first provide a general analytical finding then present some simulation-based results.

\subsection{Using the general theory of chi-squared distributions}
\label{subsection1:powerX}

One concern with the use of the proposed 3 d.f.\ test is the potential loss of power due to the increased degrees of freedom. Indeed, if the true model for an X-chromosomal SNP is without a dominance effect and skewed inactivation, without gene-sex interaction, and the true inactivation status is known so that the additive genotype variable $G_A$ can be correctly coded, then the corresponding 1 d.f.\ test will be more powerful than the proposed 3 d.f.\ test. However, we show that even under the worst-case scenario and irrespective of sample size and the nominal type I error $\alpha$ level, the maximum power loss of the proposed 3 d.f.\ is, surprisingly, capped at 18.8\%, while the potential maximum power gain is $1-\alpha$ (i.e. close to 100\%).

Let $W_1 \sim \chi^2_{(1, ncp_1)}, W_2 \sim \chi^2_{(2, ncp_2)}$ and $W_3 \sim \chi^2_{(3, ncp_3)}$ be the 1, 2 and 3 d.f.\ test statistics derived from the different regression models listed in Table \ref{summary}.  The power difference between the different $W$'s depends on both the non-centrality parameters and $\alpha$.  When all the $ncp$'s are close to 0, all tests have no power.  At the other extreme when all $ncp$'s are sufficiently large or $\alpha$ close to 1, all tests have power close to 1.  Thus, we expect meaningful power comparison when $ncp$'s, and $\alpha$ have moderate values. 

First, we assume that there are no dominance or interaction effects and the true XCI status is known to study the maximum power loss induced by unnecessarily including the $G_D$ and $GS$ terms. In that case, $ncp_1=ncp_2=ncp_3=ncp$ and $W_1$, derived from $g(E(Y))=\beta_0+\beta_S S+\beta_A G_A$ with the correct genotype coding, is the optimal test. Varying $ncp$ and $\alpha$ values, we numerically compute the power of the $W$'s for $ncp \in [0,100]$ and $-\log_{10} \alpha \in [0,15]$. Web Figure S1 provides a heat plot for power as a function of $ncp$ and $\alpha$ for the two tests.  Results show that the maximum power loss of $W_3$ compared to $W_1$ is capped at 18.8\%, regardless of the true additive effect size, sample size and the $\alpha$ level. The maximum occurs at $\alpha= 0.0008$ and $ncp = 13.4$ (Web Figure S1). At the genome-wide significance level $\alpha=5 \times 10^{-8}$ \citep{dudbridge08}, the maximum power loss is 17.7\% occurring at $ncp=32.6$. 

Notably, the maximum of 18.8\% power loss holds for comparing any 3 d.f.\ $\chi^2$ test with a 1 d.f.\ $\chi^2$ test that was derived from the {\it correctly} specified 1 d.f.\ model. This is because the derivation is based on $ncp$ and $\alpha$ alone. Second, we emphasize that although a 18.8\% loss of power is substantial, the fact that this is the maximum power loss for the 3 d.f.\ test, under any true 1 d.f.\ genetic model and regardless of the true genetic effect size, sample size and significance level, is encouraging, as the potential power gain of the proposed 3 d.f.\ test under other models can be much greater than 18.8\% as we show next. 

In the presence of dominance effect/skewed XCI or interaction effect/misspecified XCI, $ncp_3=ncp_1+\Delta_{13}$, where $\Delta_{13}>0$.  Compared to the maximum power loss of using the proposed 3 d.f.\ for a 1 d.f.\ (correctly specified) model, the maximum power gain under other genetic models can be theoretically as large as $1-\alpha$. To provide specific numerical results, we consider $\alpha=0.0008$ (the worse-case scenario derived above for the 3 d.f.\ test), $ncp_1=5$, 10 or 15, and $\Delta_{13}$ ranging from 0 to 10.  Results in Web Figure S2 show that once $\Delta_{13}$ is as large as half of $ncp_1$ (i.e.\  $ncp_3 \approx 1.5 \cdot ncp_1$), the 3 d.f.\ test is more powerful than the 1 d.f.\ test .

Together these two observations suggest that the proposed 3 d.f.\ test is not only robust to the various model uncertainties associated with analyzing X-chromosomal variants, but it is reasonably powered as compared with the standard 1 d.f.\ additive test.  Compared with a 2 d.f.\ test derived from correctly specified $g(E(Y))=\beta_0+\beta_S S+\beta_A G_A+\beta_D G_D$ or $g(E(Y))=\beta_0+\beta_S S+\beta_A G_A+\beta_{GS} G_{GS}$, the global maximum power loss of the proposed 3 d.f.\ test is capped at 7.7\%, occurring at $\alpha=9.12\times 10^{-5}$ and $ncp=19$. At $\alpha=5 \times 10^{-8}$, the maximum power loss is 7.5\% occurring at $ncp=34.2$ (Web Figure S1). In contrast, if the 2 d.f.\ model is misspecified the potential power gain of the proposed 3 d.f. test can be greater than 95\%.

Power comparison between a 1 d.f.\ test and a 2 d.f.\ test is more relevant to the analysis of an autosomal SNP, but the conclusion is similar to above. For example, under additivity, the maximum power loss of a 2 d.f.\ genotypic test is capped at 11.4\% across all parameter values and sample sizes. The maximum occurs at $\alpha=0.0025$ and $ncp=10.6$, and at the genome-wide significance level of $\alpha=5 \times 10^{-8}$ \citep{dudbridge08}, the maximum power loss is 10.3\% when $ncp=31.4$ (Web Figure S1). Web Appendix B also provides power comparison between the additive and genotypic tests for association analysis of an autosomal SNP across a range of dominance effects and allele frequencies (Web Figure S3). For each combination of parameter values considered, Web Figure S4 and Table S1 also provide the corresponding $ncp_1$ and $ncp_2$.  
  
\subsection{Using different genetic models for the X-chromosome}
\label{subsection2:powerX}

Here we provide some empirical results based on different genetic models for an X-chromosomal variant and sample sizes. Note that tests derived from models that do not include sex as a covariate are susceptible to type I error rate inflation. Thus, power comparisons here focus on $M_1$--$M_4$ as specified in Table \ref{summary}.  

To compare the empirical power, we first derive the non-centrality parameters of the tests as functions of sample size, additive, dominance and interaction effects, and under different assumptions of the baseline allele and X-inactivation status.  We provide the technical details in Web Appendices C and D. We then considered $n=1,000$, $\alpha=0.0008$ (the worst case scenario for the 3 d.f.\ test as shown in Section \ref{subsection1:powerX}), and allele frequency $f_{male}=f_{female}=0.2$ or 0.5. Results for other parameter values, including differential allele frequency values between males and females, are provided as online Supplementary Materials; sex-specific allele frequencies may occur due to sex-specific selection.

\begin{figure}

\centering \textbf{A.} $\mathbf{f_{female}=f_{male}=0.2}$
\begin{center}\includegraphics[scale=0.18]{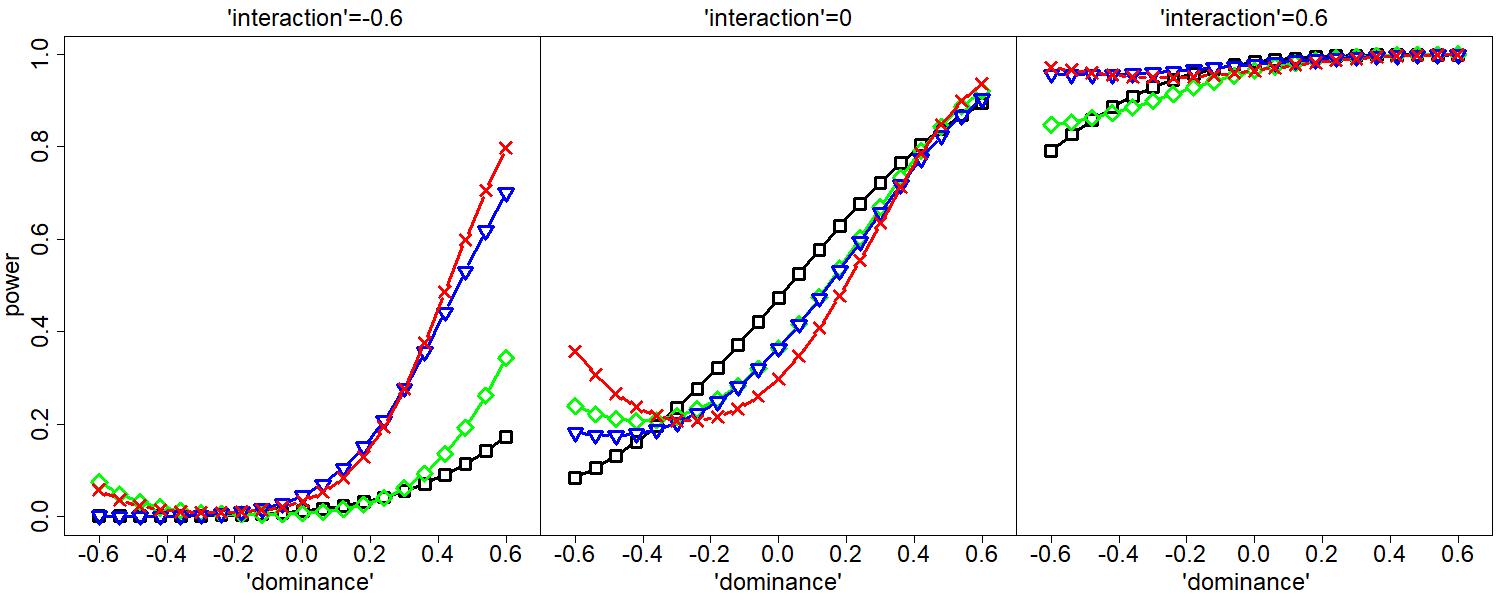}\end{center}
\begin{center}\includegraphics[scale=0.18]{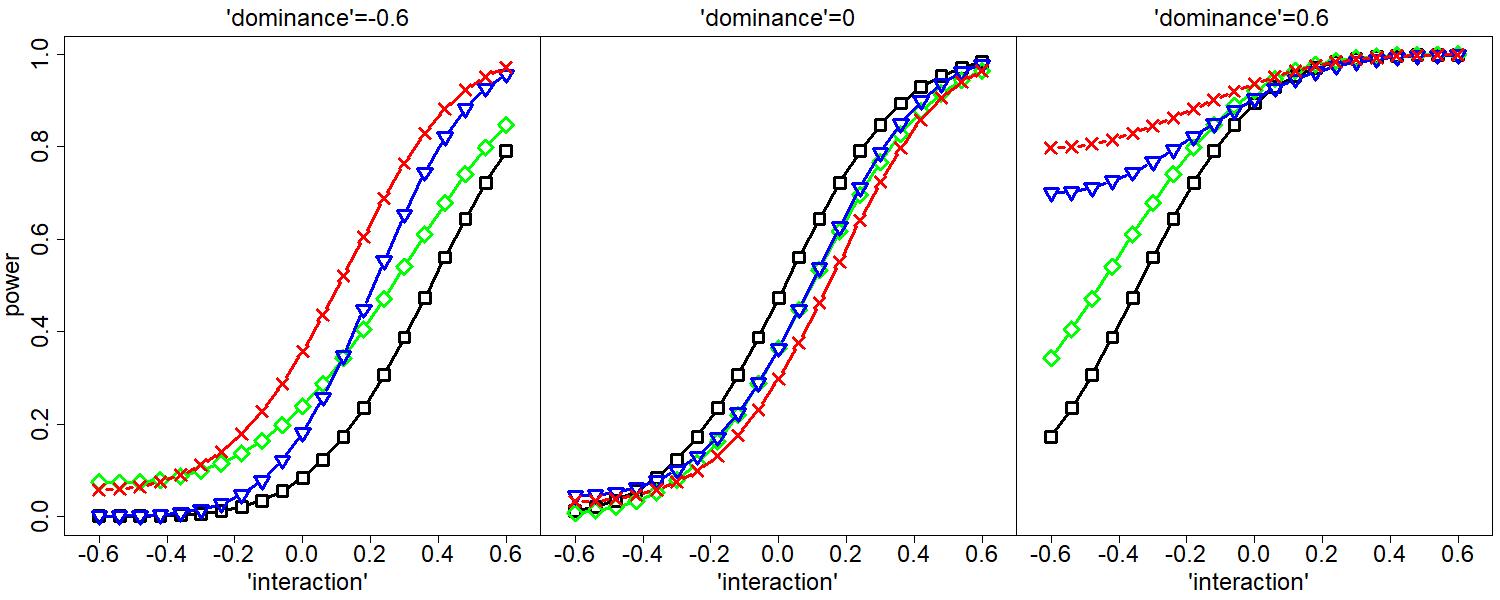}\end{center}

\centering \textbf{B.} $\mathbf{f_{female}=f_{male}=0.5}$
\begin{center}\includegraphics[scale=0.18]{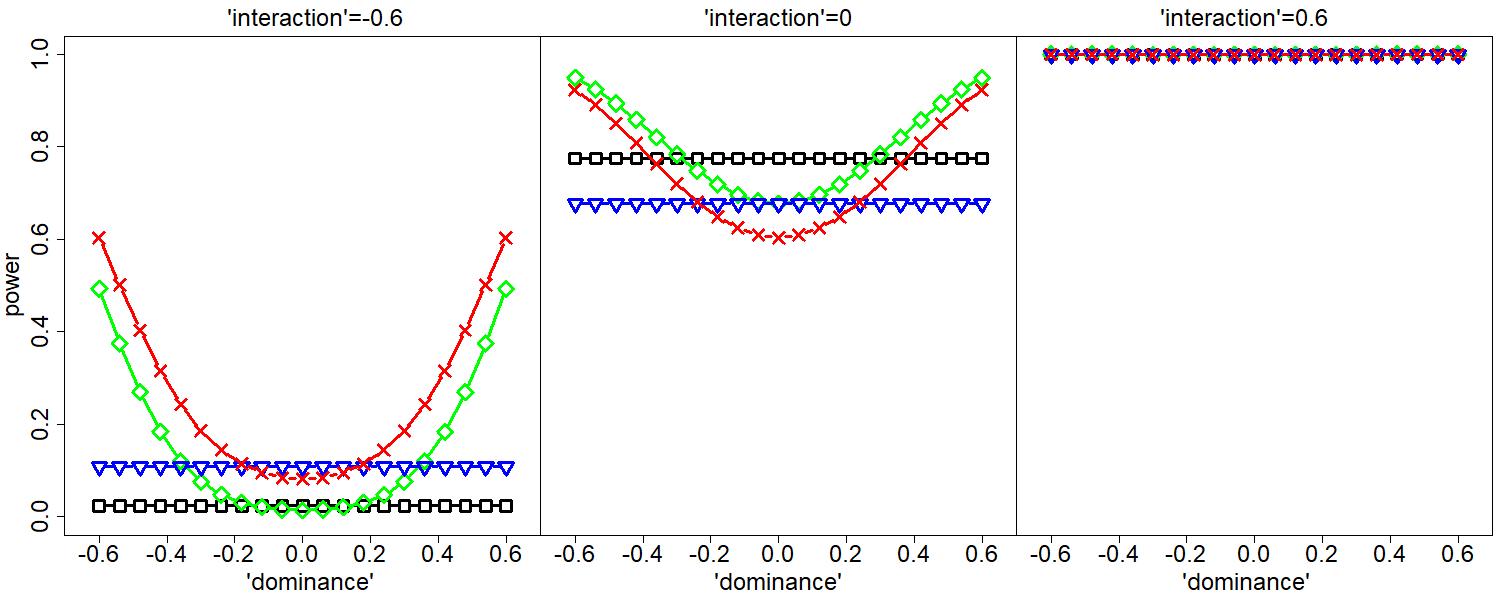}\end{center}
\begin{center}\includegraphics[scale=0.18]{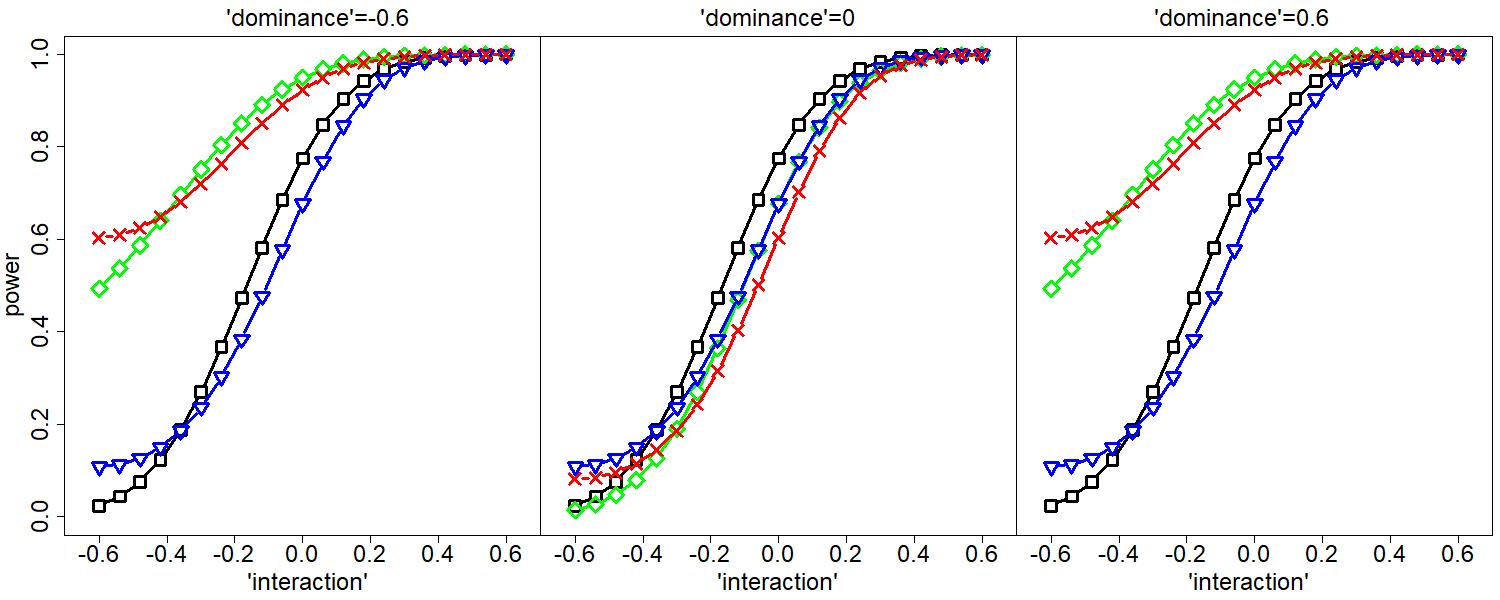}\end{center}

\caption{{\bf Power comparison for analyzing X-chromosomal SNPs.} {\color{black} Black $\square$ curves} for testing $\beta_A=0$ based on model $M_1$ as specified in Table \ref{summary}, {\color{green} green $\diamond$ curves} for testing $\beta_A=\beta_D=0$ based on model $M_2$, {\color{blue} blue $\triangledown$ curves} for testing $\beta_A=\beta_{GS}=0$ based on model $M_3$, and {\color{red} red $\times$ curves} for testing $\beta_A=\beta_D=\beta_{GS}=0$ based on the proposed model $M_4$.  \textbf{Upper panels in A and B} examine power as a function of the 'dominance' effect. \textbf{Lower panels in A and B} examine power as a function of the gene-sex `interaction' effect. Note that biological dominance effect and skewed XCI, and gene-sex interaction effect and the XCI status are statistically confounded with each other; see Section \ref{subsection2:powerX}.  Results for other parameter values including differential $f$ between males and females are shown in Web Figures S5. The analyses related to $M_1$--$M_3$ assume that the true baseline allele is known and $f$ being the allele frequency of the non-baseline allele, and the true XCI status is known at the population level. Unlike the other methods ($M_1$--$M_3$), the proposed method ($M_4$) is invariant to the assumptions of the baseline allele and the XCI status.} 
\label{powerX}  
\end{figure}

Because of the various analytical equivalencies between $GS$ interaction and XCI status, and between dominance effect and skewed XCI, the corresponding interaction, dominance and skewed effect sizes are statistically confounded with each other. Thus, we specified the averaged statistical effect size for each of the five genotype groups, i.e., $\mu_{rr}, \mu_{rR}, \mu_{RR}, \mu_r$, and $\mu_R$. We fixed $\mu_{rr}=-0.3, \mu_{RR}=0.3$ and $\mu_r=0$, and varied $\mu_{rR}$ and $\mu_R$ from $-0.6$ to $0.6$.  Note that fixing $\mu_{rr}$ and $\mu_{RR}$ is equivalent to fixing the additive effect $\beta_A=0.6$ under XCI or $\beta_A=0.3$ under no XCI; varying $\mu_{rR}$ is equivalent to varying the dominance effect $\beta_{D}$ from $-0.6$ to $0.6$.  The link with the interaction effect $\beta_{GS}$ is less clear. Under the XCI assumption, $\beta_{GS}=(\mu_R-\mu_r)-(\mu_{RR}-\mu_{rr})/2=\mu_R-0.3$, while under the no XCI assumption, $\beta_{GS}=(\mu_R-\mu_r)-(\mu_{RR}-\mu_{rr})/4=\mu_R-0.15$. Thus, for the $\mu_R$ values considered here, $\beta_{GS}$ ranged from $-0.9$ to 0.3 under XCI, and from $-0.75$ to 0.45 under no XCI.  For ease of interpretation, Figure \ref{powerX} uses the `dominance' and `interaction' terms to denote the varying degrees of $\mu_{rR}$ and $\mu_{R}$. 
 
Results in Figure \ref{powerX} demonstrate the merits of the proposed method (testing $\beta_A=\beta_D=\beta_{GS}=0$ jointly, the red $\times$ curves).  While there could be some power loss in the worse case scenario (no $G_D$ dominance or $GS$ interaction effects), it is theoretically capped at 18.8\% regardless of the parameter values.  On the other hand, compared with the standard 1 d.f.\ additive test (testing $\beta_A=0$ and assuming the correct genotype coding, the black $\square$ curves), power gain can be 70\% for the cases considered here.  When the allele frequency is $0.2$ (Figure \ref{powerX}A), the performance of the 2 d.f.\ additive and interaction test (testing $\beta_A=\beta_{GS}=0$, the blue $\triangledown$ curves) is close to the proposed 3 d.f.\ test. However, that is no longer the case when $f=0.5$ (Figure \ref{powerX}B), where the 2 d.f.\ additive and dominance test (testing $\beta_A=\beta_{D}=0$, the green $\diamond$ curves) is better and close to the proposed 3 d.f.\ test.  Web Figures S5 provides additional results for other parameter values, all showing the robustness of the proposed method, which is testing $H_0: \beta_A=\beta_D=\beta_{GS}=0$ based on $M_4$, 
$g(E(Y))=\beta_0+\beta_S S+\beta_A G_A+\beta_{D} G_D + \beta_{GS} GS$. In practice the regression model should include relevant $E$'s, which are omitted here for notation simplicity.

The proposed method not only resolves all C1--C8 analytical challenges simultaneously, but also has the best overall performance across the different underlying genetic models. However, we note that our robust method cannot identify the underlying true genetic model. This is true for any method that uses GWAS data alone, because we have shown, for example, XCI uncertainty is analytically equivalent to a gene-sex interaction effect, while XCI skewness is analytically equivalent to dominance effect. The lack of model identifiability of the proposed method, however, does not prevent a robust and powerful association analysis of X-chromosomal SNPs.

\section{Applications to Three Previously Published Association Studies}
\label{application}

\subsection{Re-analyses of the X-chromosome-inclusive genome-wide association study of \cite{sun12} }
\label{CF}

This dataset consists of 3,199 unrelated individuals with cystic fibrosis (CF) and 570,724 genome-wide bi-allelic SNPs after standard quality control \citep{sun12}. Among the 570,724 SNPs, 14,279 are for the X-chromosome and 556,445 are from the autosomes. And among the 3,199 CF subjects, 574 are cases with meconium ileus, an intestinal obstruction at birth seen in $\approx 15\%$ of CF patients \citep{dupuis16}, and the remaining 2,625 CF subjects are controls; 1,722 are males and 1,477 are females. The rates of meconium ileus are 17.7\% and 18.3\%, respectively, in the male and female groups, which is not statistically different.

A previous X-chromosome-inclusive GWAS of meconium ileus in CF has been conducted based on this dataset \citep{sun12}, where the standard 1 d.f.\ additive test was used for analyzing the autosomal SNPs, and X-chromosome being inactivated was further assumed for analyzing the X-chromosomal SNPs (i.e.\ using model $M_1$ in Table \ref{summary} with genotype coding under the assumption of XCI). Here we re-analyze both the autosomal and X-chromosomal SNPs to demonstrate the utility of the proposed approach.

\begin{figure}

\centering \textbf{A. X-chromosome Results} 
\begin{center}\includegraphics[scale=0.35]{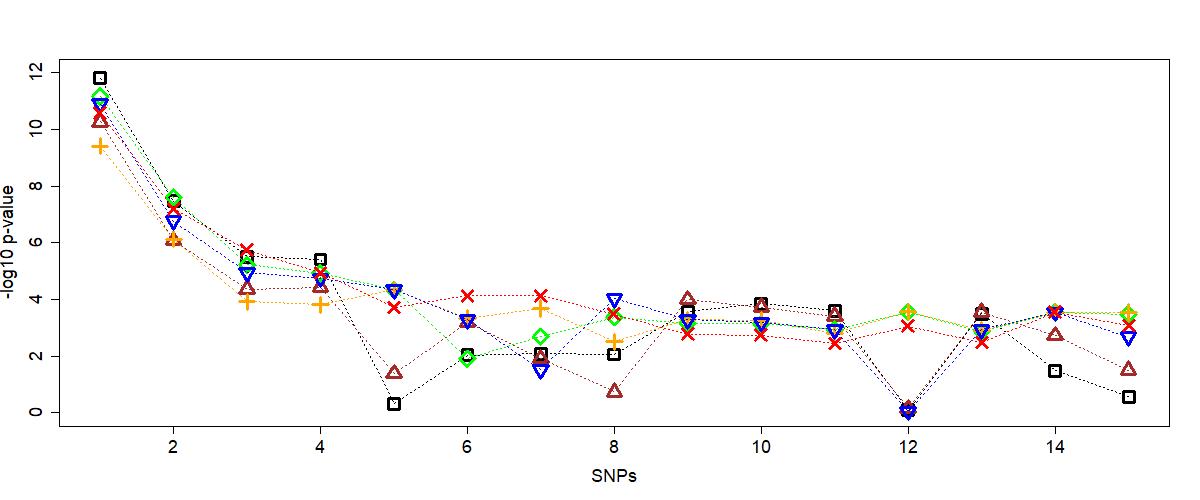}\end{center}

\centering \textbf{B. Autosome Results} 
\begin{center}\includegraphics[scale=0.35]{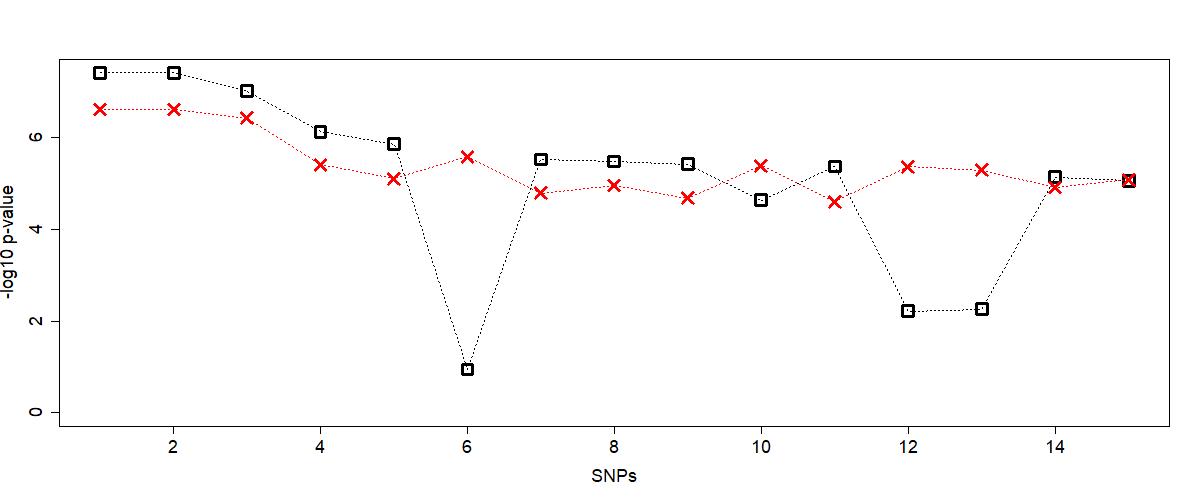}\end{center}

\caption{{\bf Results of a genome-wide association study of meconium ileus in cystic fibrosis subjects}. In total, 3,199 independent cystic fibrosis subjects, 14,279 X-chromosomal SNPs and 556,445 autosomal SNPs are analyzed. The SNPs are ordered by the minimal p-value of the different tests considered, and the lines connecting the SNPs are used only for visualization purposes to demonstrate the robustness of a particular method. 
{\bf A}: These top 15 ranked X-chromosomal SNPs are selected based on any of the six tests based on $M_1$--$M_4$ models in Table \ref{summary}: the {\color{black} Black $\square$ curve} for testing $\beta_A=0$ based on $M_1$ assuming X-chromosome inactivation (XCI), the {\color{brown} brown $\triangle$ curve} for testing $\beta_A=0$ based on $M_1$ assuming no XCI, the {\color{green} green $\diamond$ curve} for testing $\beta_A=\beta_D=0$ based on $M_2$ assuming XCI, the {\color{orange} orange $+$ curve} for testing $\beta_A=\beta_D=0$ based on $M_2$ assuming no XCI, the {\color{blue} blue $\triangledown$ curve} for testing $\beta_A=\beta_{GS}=0$ based on $M_3$ (invariant to the XCI assumptions if $GS$ is included in the model and tested), and the {\color{red} red $\times$ curve} for testing $\beta_A=\beta_D=\beta_{GS}=0$ based on the recommended model $M_4$ that is most robust for analyzing the X-chromosome.
{\bf B}: These top 15 ranked autosomal SNPs are selected based on either the 1 d.f.\ additive test or the 2 d.f.\ genotypic test. The {\color{black} black $\square$ curve} for testing $\beta_A=0$ using the standard additive model, and the {\color{red} red $\times$ curve} for testing $\beta_A=\beta_D=0$ using the recommend genotypic model that is most robust for analyzing the autosomes.}
\label{top15}  
\end{figure}

For the X-chromosome, we compared the $M_1$--$M_4$ models and their corresponding tests as detailed in Table \ref{summary} in Section \ref{Xchr}.  For each SNP, we performed six different association tests, depending on which of the $M_1$--$M_4$ models was used and if the XCI status needed to be specified, because (a) sex must be included to ensure correct type I error rate control and models including $S$ are invariant to the choice of the baseline allele (Section \ref{includingS}), and (b) models including the gene-sex interaction effect are invariant to the assumption of XCI (Section \ref{includingGS}). Figure \ref{top15}A shows the results for the top 15 ranked X-chromosomal SNPs, ordered by the minimal p-value of all six tests; the lines connecting the SNPs are used only for visualization purposes to demonstrate the robustness of a method.

The application results here are consistent with our earlier analytical and simulation results in Section \ref{section:powerX}, showing that joint modelling and testing the additive, dominance and gene-sex interaction effects is the most robust association approach for analyzing X-chromosomal variants. For example, the result of  \cite{sun12} ($M_1$ and assuming XCI) marked by the black $\square$ curve is clearly less `stable' than the red $\times$ curve (the proposed $M_4$) across different SNPs. In this particular application, we observed that the performance of the  orange $+$ curve ($M_2$ assuming no XCI) is similar to the proposed method. However, interestingly, the green $\diamond$ curve (also $M_2$ but assuming XCI) is noticeably different from the the  orange $+$ curve. 

For the autosomal SNPs, we contrast the standard 1 d.f.\ additive test with the proposed 2 d.f.\ genotypic test as briefly discussed in section \ref{subsection1:powerX} and detailed in Web Appendix B.  Figure \ref{top15}B shows the results for the top 15 ranked autosomal SNPs, ordered by the minimal p-value of additive and genotypic tests; Web Figures S7 and S8  provide genome-wide results. It is clear that if the p-values of the standard 1 d.f.\ additive test (the black $\square$ curve) are smaller, then those from the recommended 2 d.f.\ genotypic test (the red $\times$ curve) are close in magnitude, while the reverse is not true. For example, p-value of the recommended 2 d.f.\ genotypic test for the $6_{th}$ SNP ($rs2657147$) in the plot is more than four orders of magnitude smaller than that of the 1 d.f.\ additive test; there is no evidence for genotyping error at this SNP as the p-value of HWE test in the control group is $0.026$. The genotype counts for $rr$, $rR$, and $RR$ are $(210, 312, 52)$ in the case group and $(1012, 1192, 421)$ in the control group, which yields case/control ratios of $(0.208, 0.262, 0.124)$, clearly suggesting a dominance pattern. Whether this is a true new finding, however, requires further investigation.

\subsection{Evidence from the first (autosome only) genome-wide association study of \cite{WTCCC}} 
\label{WTCCC}

We then examined the results of the first (autosome only) genome-wide association study, conducted by the Wellcome Trust Case Control Consortium \citep{WTCCC}. Their Table 3 lists regions of the genome showing the strongest association signals and provides results from both the 1 d.f.\ trend test (statistically equivalent to the additive test considered here) and the 2 d.f.\ genotypic tests. 

Consistent with the autosomal results of the CF meconium ileus application above, the results in Table 3 of  \cite{WTCCC} also show that if the 1 d.f.\ additive test provides a smaller p-value, the p-value of the 2 d.f.\ genotypic test is at most one order of magnitude larger. For example, the p-values are $1.16\times 10^{-13}$ and $1.79\times 10^{-14}$, respectively, for the 1 d.f.\ additive and  2 d.f.\ genotypic tests, testing association between coronary artery disease and rs1333049, the second SNP in Table 3 of \cite{WTCCC}. On the other hand, the p-value of the 2 d.f.\ genotypic test can be several orders of magnitude smaller that of the 1 d.f.\ additive test. For example, the p-values are \ $2.19\times 10^{-4}$  and $6.29\times 10^{-8}$, respectively, for the 1 d.f.\ additive and  2 d.f.\ genotypic tests, testing association between bipolar disorder and rs420259, the first SNP in Table 3 of \cite{WTCCC}; the association between rs420259 and bipolar disorder has since been replicated by other studies \citep{tesli10, gonzalez16}. We can draw similar conclusions based on the Bayes factors provided in their Table 3, obtained under the 1 d.f.\ additive or 2 d.f.\ genotypic models.

\subsection{Re-analyses of the 60 autosomal SNPs potentially associated with various complex traits, selected by \cite{wt05}}
\label{60SNPs}

Finally, we re-analyzed the 60 autosomal SNPs selected by \cite{wt05} from 41 case-control association studies of various complex traits, including Alzheimer disease and breast cancer; the genotype count data are available from Table 1 of \cite{wt05}. Although these SNPs were originally selected by \cite{wt05} for a study of departure from Hardy-Weinberg equilibrium, genotype-based methods are robust to the HWE assumption \citep{sasieni97, zhang21}.  Here we focused on comparing the standard 1 d.f.\ additive test with the recommended 2 d.f.\ genotypic test for analyzing these 60 autosomal SNPs, which are presumed to be associated with complex traits based on the earlier 41 studies.  

We observed that the genotypic test leads to 31 SNPs with p-values less than $\alpha=0.05$, while the additive test results in 22 SNPs (Web Figure S6). Using the Bonferroni threshold of $\alpha=0.05/60$ the numbers are 7 and 6, respectively, for the genotypic and additive tests.  Although these autosomal SNPs can only be presumed to be associated with the various complex traits, the empirical evidence here is consistent with the analytical and simulation results in section \ref{subsection1:powerX} and Web Appendix B.

\section{Discussion}
\label{discussion}

We have shown that in association analysis of an X-chromosomal variant, the sex main effect must be included to achieve correct type I error rate control.  The inclusion of sex also addresses the complication of baseline allele specification that otherwise affects association inference for an X-chromosomal SNP, in contrast to an autosomal SNP.  Although the method developed here is motivated by genetic association studies of the X-chromosome, Theorem \ref{transformation} is applicable to other settings where model uncertainty plays a role. For association studies of autosomal variants, sex is not routinely included. However, sex can be a confounder for an autosomal SNP as well, e.g.\ when there is sex difference in allele frequency due to sex-specific selection. When the allele frequency difference is small, including sex does not substantively change the association result, because sex is not directly included in the genetic association test. Thus, we recommend to always include sex as a covariate in association analysis of either autosomal or X-chromosomal variants. 

We have also shown that modelling the genetic dominance effect $\beta_D$ is beneficial for analyzing both the autosomes and X-chromosome. The proposed model can significantly increase test power when $\beta_D$ is large. When $\beta_D$ is close to 0, the model is still robust and maintains `comparable' power with that of the additive model; `comparable' is in the context of the trade-off between the maximum power loss and gain across different models. For an autosomal SNP, we have shown analytically that even under true additivity, compared with the classical 1 d.f. additive test, the maximum power loss of the 2 d.f. genotypic test is capped at 11.4\%, regardless of the sample, genetic effect and test sizes, but the power gain can be as high as $1-\alpha$. Similarly, for an X-chromosome SNP, with a 3 d.f.\ test that includes $\beta_A$, $\beta_D$ and $\beta_{GS}$ interaction effects, power loss is capped at 18.8\%; this assumes that the standard 1 d.f.\ additive test used the correct XCI model and there is no skewed XCI or dominance effect. If these assumptions do not hold, the potential power gain of the 3 d.f.\ test can be as high as $1-\alpha$. However, not all alternative genetic models are equally likely in practice. Consistent with the earlier work of \cite{hill08} and \cite{bush12}, two recent studies showed that ``genetic variance for complex traits is predominantly additive" \citep{hivert21,pazokitoroudi21}. To this end, a Bayesian alternative that incorporates prior evidence for the different genetic models can be considered.  

When the true genetic model is unknown, one alternative frequentist's approach is to consider all possible models and use the `best' or weighted average.  But, such an approach is difficult to implement in practice; see \cite{bagos13} for a review.  For example, selection bias inherent in choosing the best-fitted model must be corrected for, often through computationally intensive simulation studies, and power of this bias-corrected inferential procedure is not clear. On the other hand, ways to obtain a weighted average of the test statistics or p-values across all models can be quite ad hoc, and the optimal weighting factors are difficult to derive. The recent Cauchy method can be used to combine correlated p-values derived from all possible genetic models \citep{liu20}. Finally, the method proposed here is tailored for analyzing one common SNP at a time, and joint analysis of multiple common or rare SNPs \citep{andriy14} requires further consideration. 

When the true genetic model is unknown, another alternative is to use sex-stratified analysis, followed by meta-analysis combining the female and male groups \citep{willer10}. This approach appears to be robust to the XCI assumption when analyzing an X-chromosomal variant, because association evidence in females are the same between the XCI and no-XCI assumptions. However, the two assumptions lead to different effect size estimates by a factor of two (the standard errors also differ by a factor of two), resulting in different results using the inverse-variance based meta-analysis. The sample-size based meta-analysis can overcome this limitation, but other issues remain including difficulty of modelling non-additive or gene-sex interaction effects. 

Summary statistics from the proposed 3 d.f.\ test for X-chromosome SNPs can be used to perform meta-analysis by using, for example, Fisher's combined p-value approach. The classical inverse-variance-based method, however, is not applicable for two reasons. Firstly, there are multiple genotype-related $\beta$ estimates, $\beta_A$, $\beta_D$ and $\beta_{GS}$. Secondly, and more importantly, some of the $\beta$ estimates are not meaningful on their own as we have shown that skewed XCI and dominance effects are statistically confounded with each other, so are the $G\times S$ interaction effect and the assumption of XCI. Even if we limit our attention to the genetic main additive effect, the effect size estimate changes by a factor of two depending on the XCI assumption (i.e. the genotype coding scheme). Thus, our work here also highlights new challenges associated with other analyses of X-chromosomal SNPs. For example, how to aggregate association evidence across multiple SNPs or multiple traits \citep{zhao21}, and how to perform X-chromosome-inclusive polygenic risk score (PRS) analysis \citep{dudbridge18}, both of which we will address in future research.
 
The proposed full model for analyzing an X-chromosomal SNP, $g(E(Y))=\beta_0+\beta_SS+\beta_A G_A + \beta_D G_D+\beta_{GS} GS$, is robust to various model uncertainties,  analytically. However, as noted earlier it is not capable of differentiating between the scenarios. Using the available genetic association data, \cite{ma15} proposed a variance-based test for detecting X-inactivation by comparing phenotypic variance of the $rR$ group with that of the $rr$ and $RR$ groups in females, but this method is limited to a continuous trait \citep{soave17,deng19}. \cite{wang14} explicitly introduced a parameter to represent the amount of skewness of X-inactivation. Our work here, however, shows that the interpretation of their parameter is statistically confounded with dominance genetic effect using GWAS data alone. How to incorporate additional `omic' data  \citep{carrel05} to tease apart different biological phenomena is an interesting problem that deserves further investigation. 


\section*{Acknowledgments}
The authors thank the two referees for their helpful and constructive comments, which have substantially improved the presentation of the paper. The authors thank cystic fibrosis patients and their families who participate in the International CF Gene Modifier studies, the US and Canadian CF Foundations for the genotyping and clinical data of the International CF Gene Modifier Consortium, and the members of the consortium Johanna Rommens, Garry Cutting, Michael Knowles, Mitchell Drumm and Harriet Corvol. The authors would also like to thank Prof. Keith Knight and Prof. Nancy Reid for their helpful comments, and Mr. Bowei Xiao for assisting the CF application.  

\subsection*{Author contributions}

Bo Chen developed the method and designed the study with significant input from Lei Sun and Radu Craiu. Bo Chen performed the data analysis with significant input from Lei Sun and Lisa Strug. Bo Chen wrote the manuscript with significant contributions from all co-authors. All authors have reviewed and approved the final manuscript.

\subsection*{Financial disclosure}

This research is funded by the Natural Sciences and Engineering Research Council of Canada (NSERC) to RVC (RGPIN-249547) and LS (RGPIN-04934; RGPAS-522594), the Canadian Institutes of Health Research (CIHR) to LS (MOP-310732) and LJS (MOP-258916), and the Cystic Fibrosis Canada to LJS (2626). 

\subsection*{Conflict of interest}

The authors declare no potential conflict of interests.

\section*{Data availability statement}

The meconium ileus application data is available by application to the Cystic Fibrosis Canada National data registry for researchers who meet the criteria for access to confidential clinical data for the purpose of CF research. The 60-SNP application data is publicly available from Table 1 of \cite{wt05}. The WTCCC application used only the summary statistics reported in Table 3 of  \cite{WTCCC}.

\section*{Supporting information}

The online supplementary materials contain proof of Theorem 1 in Appendix A, additional results of the power study in Appendix B, computations of the non-centrality parameters for correctly specified genetic models in Appendix C and for misspecified genetic models in Appendix D, and additional figures in Appendix E.

\bibliography{mybib}%

\end{document}


\def\spacingset#1{\renewcommand{\baselinestretch}%
{#1}\small\normalsize} \spacingset{1}


\if1\blind
{
  \title{\bf Supplementary Materials for The X factor: A Robust and Powerful Approach to X-chromosome-Inclusive Whole-genome Association Studies}
  \author{Bo Chen, Radu V. Craiu, Lisa J. Strug and Lei Sun \hspace{.2cm}}
  \date{}
  \maketitle

} \fi

\if0\blind
{
  \bigskip
  \bigskip
  \bigskip
  \begin{center}
    {\LARGE\bf Supplementary Materials for The X factor: A Robust and Powerful Approach to X-chromosome-Inclusive Whole-genome Association Studies}
\end{center}
  \medskip
} \fi

\newpage
\spacingset{1.5} 
\section*{Appendix A: Proof of Theorem 1}
In Appendix A, we prove Theorem 1 stated in Section 2.2. 

\noindent{\bf Case A: Linear Regression} \\ We start from the special case of linear model. For notation simplicity we first rewrite the null hypotheses in matrix form: $H_0: L\boldsymbol{\beta_1}=0$ under $\mathcal{M}_1$ and $H_0: L\boldsymbol{\beta_2}=0$ under $\mathcal{M}_2$, where $L=(0_{(q, p-q)}, I_q)$ is a combination of $q \times (p-q)$ zero matrix and identity matrix with dimension $q$. 

In the case of the linear model, it is well known \citep{vandaele81} that the Wald, Score and LRT test statistics for $H_0$ are all functions of the F-statistic. So it is sufficient to show that the F-statistic is the same for  $\mathcal{M}_1$ and $\mathcal{M}_2$. Specifically, the F-statistic under $\mathcal{M}_j$ for $j=1, 2$ is
$$F_j=\frac{Q_j/q}{(Y-X_j(X_j'X_j)^{-1}X_j'Y)'(Y-X_j(X_j'X_j)^{-1}X_j'Y)/(n-p)} \sim F(q,n-p),$$
where
$$Q_j=Y'X_j(X_j'X_j)^{-1}L'(L(X_j'X_j)^{-1}L')^{-1}L(X_j'X_j)^{-1}X_j'Y.$$
If $X_j$ denotes the covariate  matrix used in model $\mathcal{M}_j$, then consider a  partition of its columns into $X_j=(X_{j1}  X_{j2})$ such that the effect of $X_{jk}$ on the response is $\beta_{jk}$  for  $j,k=1,2$. We partition $(X_j'X_j)^{-1}$ into 4 blocks: $(X_j'X_j)^{-1}=\left(\begin{array}{cc}
                   X_j^{11} & X_j^{12} \\
                   X_j^{21} & X_j^{22} \\
           \end{array}\right)$.
Then $L'(L(X_1'X_1)^{-1}L')^{-1}L$ is simplified to $\left(\begin{array}{cc}
                   0 & 0 \\
                   0 & (X_1^{22})^{-1} \\
           \end{array}\right)$,
which implies 
$$Q_1=Y'X_1(X_1'X_1)^{-1}\left(\begin{array}{cc}
                   0 & 0 \\
                   0 & (X_1^{22})^{-1} \\
           \end{array}\right)(X_1'X_1)^{-1}X_1'Y$$
Next, $X_2=X_1T$ implies $L'(L(X_2'X_2)^{-1}L')^{-1}L=\left(\begin{array}{cc}
                   0 & 0 \\
                   0 & T_{22}'(X_1^{22})^{-1}T_{22} \\
           \end{array}\right),$
and
\begin{eqnarray*}
&&(T')^{-1}L'(L(X_2'X_2)^{-1}L')^{-1}LT^{-1} \\
&=&\left(\begin{array}{cc}
                   (T_{11}')^{-1} & 0 \\
                   (T_{22}')^{-1}T_{12}'(T_{11}')^{-1} & (T_{22}')^{-1} \\
           \end{array}\right)\left(\begin{array}{cc}
                   0 & 0 \\
                   0 & T_{22}'(X_1^{22})^{-1}T_{22} \\
           \end{array}\right)\left(\begin{array}{cc}
                   T_{11}^{-1} & T_{11}^{-1}T_{12}T_{22}^{-1} \\
                   0 & T_{22}^{-1} \\
           \end{array}\right) \\
&=&\left(\begin{array}{cc}
                   0 & 0 \\
                   0 & (X_1^{22})^{-1} \\
           \end{array}\right).
\end{eqnarray*}
Hence,
\begin{eqnarray*}
Q_2&=&Y'X_1TT^{-1}(X_1'X_1)^{-1}(T')^{-1}L'(L(X_2'X_2)^{-1}L')^{-1}LT^{-1}(X_1'X_1)^{-1}(T')^{-1}T'X_1'Y \\
&=&Y'X_1(X_1'X_1)^{-1}\left(\begin{array}{cc}
                   0 & 0 \\
                   0 & (X_1^{22})^{-1} \\
           \end{array}\right)(X_1'X_1)^{-1}X_1'Y=Q_1.
\end{eqnarray*}
On the other hand, $X_2(X_2'X_2)^{-1}X_2=X_1TT^{-1}(X_1'X_1)^{-1}(T')^{-1}T'X_1'=X_1(X_1'X_1)^{-1}X_1$. Therefore, $F_1=F_2$. Finally, the Wald, Score and LRT statistics are
$$Wald=\frac{nqF}{n-p}; Score=\frac{nqF}{qF+n+p}; LRT=n\log(1+\frac{qF}{n-p}).$$
Since $F$ does not change, they are all invariant to the linear transformation $T$ between $\mathcal{M}_1$ and $\mathcal{M}_2$.

\noindent{\bf Case B: Generalized Linear Regression.} \\
In generalized linear model, the three test statistics usually do not have closed forms, and they are calculated from $\boldsymbol{\hat\beta_j}$ and $\boldsymbol{\tilde{\beta_j}}$, the unconstrained and constrained MLE of $\boldsymbol{\beta_j}$, which are usually estimated by numerical methods. Throughout the proof, we use $\hat\cdot$ and $\tilde\cdot$ to denote unconstrained and, respectively, constrained (under $H_0$) estimators.  For sample size $n$, we use the standard notations of GLM, where 
$$\boldsymbol{\mu}=(\mu_{1},...,\mu_{n})=g^{-1}(X\boldsymbol{\beta}),$$ 
$$V(\mu_i)=Var(Y_i)/\phi, V(\boldsymbol{\mu})=diag[V(\mu_1),...,V(\mu_n)],$$ 
$$w(\mu_i)=1/(V(\mu_i)[g'(\mu_i)]^2), W(\boldsymbol{\mu})=diag[w(\mu_1),...,w(\mu_n)],$$ 
$$z(\mu_i)=g^{-1}(\mu_i)+g'(\mu_i)(Y_i-\mu_i), z(\boldsymbol{\mu})=[z(\mu_1),...,z(\mu_n)].$$ 
The proof under generalized linear model relies on the following two assumptions:
\begin{enumerate}
\item $\boldsymbol{\hat{\beta_j}}$ and $\boldsymbol{\tilde{\beta_j}}$ are estimated by iterative reweighted least squares method, and they have an initial value of 0, i.e., $\boldsymbol{\hat{\beta_j}}^{(0)}=\boldsymbol{\tilde{\beta_j}}^{(0)}=\mathbf{0}$.
\item If the dispersion parameter $\phi$ is unknown, it is estimated using $\hat{\phi}=h(\boldsymbol{\hat{\mu}})$ and $\tilde{\phi}=h(\boldsymbol{\tilde{\mu}})$ for some function $h$.
\end{enumerate}

These two assumptions are commonly satisfied in GLM framework. For assumption 1, there is no prior information that the effect size is positive or negative, so it is  reasonable to choose an initial value of zero.  For assumption 2, there exist several possible estimators of $\phi$ in practice, but the most commonly used estimators are all functions of $\mu$, e.g., $\hat{\phi}=\frac{1}{n}\sum_{i=1}^n \frac{(Y_i-\hat{\mu_i})^2}{V(\hat{\mu_i})}$. 

We first show that  $X_1\boldsymbol{\hat{\beta_1}}=X_2\boldsymbol{\hat{\beta_2}}$, and $X_1\boldsymbol{\tilde{\beta_1}}=X_2\boldsymbol{\tilde{\beta_2}}$. Under assumption 1, $X_1\boldsymbol{\hat{\beta_1}}^{(0)}=X_2\boldsymbol{\hat{\beta_2}}^{(0)}$. If we further assume $X_1\boldsymbol{\hat{\beta_1}}^{(k)}=X_2\boldsymbol{\hat{\beta_2}}^{(k)}$, then it yields $\boldsymbol{\hat{\mu_1}}^{(k)}=\boldsymbol{\hat{\mu_2}}^{(k)}$, $V(\boldsymbol{\hat{\mu_1}}^{(k)})=V(\boldsymbol{\hat{\mu_2}}^{(k)})$, $W(\boldsymbol{\hat{\mu_1}}^{(k)})=W(\boldsymbol{\hat{\mu_2}}^{(k)})$ and $z(\boldsymbol{\hat{\mu_1}}^{(k)})=z(\boldsymbol{\hat{\mu_2}}^{(k)})$. At ($k+1$)th iteration, 
\begin{eqnarray*}
X_2\boldsymbol{\hat{\beta_2}}^{(k+1)}&=&X_2[X_2'W(\boldsymbol{\hat{\mu_2}}^{(k)})X_2]^{-1}
X_2'W(\boldsymbol{\hat{\mu_2}}^{(k)})z(\boldsymbol{\hat{\mu_2}}^{(k)}) \\
&=& X_1TT^{-1}[X_1'W(\boldsymbol{\hat{\mu_1}}^{(k)})X_1]^{-1}
(T')^{-1}T'X_1'W(\boldsymbol{\hat{\mu_1}}^{(k)})z(\boldsymbol{\hat{\mu_1}}^{(k)}) \\
&=& X_1\boldsymbol{\hat{\beta_1}}^{(k+1)}.
\end{eqnarray*}
Therefore, mathematical induction and a simple limiting argument lead to  $X_1\boldsymbol{\hat{\beta_1}}=X_2\boldsymbol{\hat{\beta_2}}$. Under the null hypothesis, we use the same argument on the submatrix $X_{11}$, $X_{21}$ and transformation matrix $T_{11}$ to show $X_{11}\tilde{\beta_{11}}=X_{21}\tilde{\beta_{21}}$, which leads to $X_{1}\boldsymbol{\tilde{\beta_{1}}}=X_{2}\boldsymbol{\tilde{\beta_{2}}}$. It immediately follows that $\boldsymbol{\hat{\beta_1}}=T\boldsymbol{\hat{\beta_2}}$, $\boldsymbol{\tilde{\beta_1}}=T\boldsymbol{\tilde{\beta_2}}$, $\boldsymbol{\hat{\mu_1}}=\boldsymbol{\hat{\mu_2}}$ and $\boldsymbol{\tilde{\mu_1}}=\boldsymbol{\tilde{\mu_2}}$.

Depending on the type of GLM, the dispersion parameter is either known (e.g., $\phi=1$ in logistic model) or unknown (e.g., $\phi=\sigma^2$ in linear model). However, the estimators of $\beta$ remain the same regardless of whether the dispersion parameter $\phi$ is known or unknown. When $\phi$ is known, it is trivial that $\phi$ remains equal under $\mathcal{M}_1$ and $\mathcal{M}_2$. When $\phi$ is unknown, we  replace $\phi$ by its estimator. Because $\boldsymbol{\hat{\mu_1}}=\boldsymbol{\hat{\mu_2}}$ and $\boldsymbol{\tilde{\mu_1}}=\boldsymbol{\tilde{\mu_2}}$, under assumption 2, $\hat{\phi}$ and $\tilde{\phi}$ also remain unchanged under $\mathcal{M}_1$ and $\mathcal{M}_2$. We discuss below each of the three tests, Wald, Score and LRT in detail.\\

 \textbf{(i) The Wald statistic} is
$$Wald_j=\frac{n}{\hat{\phi}}\{\boldsymbol{\hat{\beta_j}}'L'[L(X_j'W(\hat{\mu_j})X_j)^{-1}L']^{-1}L\boldsymbol{\hat{\beta_j}}\}.$$
Because $\boldsymbol{\hat{\beta_2}}=T^{-1}\boldsymbol{\hat{\beta_1}}$ and $W(\boldsymbol{\hat{\mu_2}})=W(\boldsymbol{\hat{\mu_1}})$, 
$$Wald_2=\frac{n}{\hat{\phi}}(\boldsymbol{\hat{\beta_1}}'(T^{-1})'L'[LT^{-1}(X_1'W(\boldsymbol{\hat{\mu_1}})X_1)^{-1}(T')^{-1}L']^{-1}LT^{-1}\boldsymbol{\hat{\beta_1}}).$$
We consider a partition $(X_1'W(\boldsymbol{\hat{\mu_1}})X_1)^{-1}$ and follow the approach used for {\bf Case A} to show
$$(T^{-1})'L'[LT^{-1}(X_1'W(\boldsymbol{\hat{\mu_1}})X_1)^{-1}(T')^{-1}L']^{-1}LT^{-1}=L'[L(X_1'W(\boldsymbol{\hat{\mu_1}})X_1)^{-1}L']^{-1}L.$$
Therefore, $Wald_1=Wald_2$. \\

\textbf{(ii) The Score statistic }is defined by \citet{cordeiro93} as
$$Score_j=\frac{1}{\tilde{\phi}}(Y-\boldsymbol{\tilde{\mu_j}})'V(\boldsymbol{\tilde{\mu_j}})^{-1/2}W(\boldsymbol{\tilde{\mu_j}})^{1/2}X_{j2}(\tilde{R_j}'W(\boldsymbol{\tilde{\mu}})\tilde{R_j})^{-1}X_{j2}'W(\boldsymbol{\tilde{\mu_j}})^{1/2}
V(\boldsymbol{\tilde{\mu_j}})^{-1/2}(Y-\boldsymbol{\tilde{\mu_j}}),$$
where
$$R_j=X_{j2}-X_{j1}(X_{j1}'W(\boldsymbol{\mu_j})X_{j1})^{-1}X_{j1}'W(\boldsymbol{\mu_j})X_{j2}.$$ 
First, we note that
$$(X_{21}, X_{22})=(X_{11}, X_{12})\left(\begin{array}{cc}
                   T_{11} & T_{12} \\
                   0 & T_{22} \\
           \end{array}\right)
=(X_{11}T_{11}, X_{11}T_{12}+X_{12}T_{22})$$
and $W(\boldsymbol{\tilde{\mu_2}})=W(\boldsymbol{\tilde{\mu_1}})$. Thus
\begin{eqnarray*}
\tilde{R_2}&=&X_{11}T_{12}+X_{12}T_{22}-X_{11}T_{11}(T_{11}'X_{11}'W(\boldsymbol{\tilde{\mu_2}})X_{11}T_{11})^{-1}T_{11}'X_{11}'W(\boldsymbol{\tilde{\mu_2}})(X_{11}T_{12}+X_{12}T_{22}) \\
&=& X_{12}T_{22}-X_{11}(X_{11}'W(\boldsymbol{\tilde{\mu_1}})X_{11})^{-1}X_{11}'W(\boldsymbol{\tilde{\mu_2}})X_{12}T_{22})=\tilde{R_1}T_{22}
\end{eqnarray*}
From the estimating equations for the constrained MLE of $\boldsymbol{\beta}$, $$(Y-\boldsymbol{\tilde{\mu_j}})'V(\boldsymbol{\tilde{\mu_j}})^{-1/2}W(\boldsymbol{\tilde{\mu_j}})^{1/2}X_{j1}=0.$$ 
Hence,
\begin{eqnarray*}
(Y-\boldsymbol{\tilde{\mu_2}})'V(\boldsymbol{\tilde{\mu_2}})^{-1/2}W(\boldsymbol{\tilde{\mu_2}})^{1/2}X_{22}&=&
(Y-\boldsymbol{\tilde{\mu_1}})'V(\boldsymbol{\tilde{\mu_1}})^{-1/2}W(\boldsymbol{\tilde{\mu_1}})^{1/2}(X_{11}T_{12}+X_{12}T_{22}) \\
&=&(Y-\boldsymbol{\tilde{\mu_1}})'V(\boldsymbol{\tilde{\mu_1}})^{-1/2}W(\boldsymbol{\tilde{\mu_1}})^{1/2}X_{12}T_{22}.
\end{eqnarray*}
Therefore,
$$Score_2=\frac{1}{\tilde{\phi}}(Y-\boldsymbol{\tilde{\mu_1}})'V(\boldsymbol{\tilde{\mu_1}})^{-1/2}W(\boldsymbol{\tilde{\mu_1}})^{1/2}X_{12}T_{22}
(T_{22}'\tilde{R_1}'W(\boldsymbol{\tilde{\mu_1}})\tilde{R_1}T_{22})^{-1} \cdotp$$
$$T_{22}'X_{12}'W(\boldsymbol{\tilde{\mu_1}})^{1/2}
V(\boldsymbol{\tilde{\mu_1}})^{-1/2}(Y-\boldsymbol{\tilde{\mu_1}}) =Score_1.$$

 \textbf{(iii) The LRT statistic} is
$$LRT_j=2\sum_{i=1}^n[\log f(Y_i, \boldsymbol{\hat{\beta_j}})-\log f(Y_i,\boldsymbol{\tilde{\beta_j}})].$$
The density function of $Y_i$ belongs to the exponential family
$$f(Y_i, \boldsymbol{\beta_j})=\exp\left [\frac{Y_iX_{ij}'\boldsymbol{\beta_j}-b(X_{ij}'\boldsymbol{\beta_j})}{\phi}+c(Y_i, \phi)\right],$$ 
so $X_1\boldsymbol{\hat{\beta_1}}=X_2\boldsymbol{\hat{\beta_2}}$ and $X_1\boldsymbol{\tilde{\beta_1}}=X_2\boldsymbol{\tilde{\beta_2}}$ imply $f(Y_i, \boldsymbol{\hat{\beta_1}})=f(Y_i, \boldsymbol{\hat{\beta_2}})$ and $f(Y_i, \boldsymbol{\tilde{\beta_1}})=f(Y_i, \boldsymbol{\tilde{\beta_2}})$. Therefore, $LRT_1=LRT_2$. 

\endproof

\section*{Appendix B: Additional Results of The Power Study}

\begin{figure}
 \centerline{\includegraphics[scale=0.25]{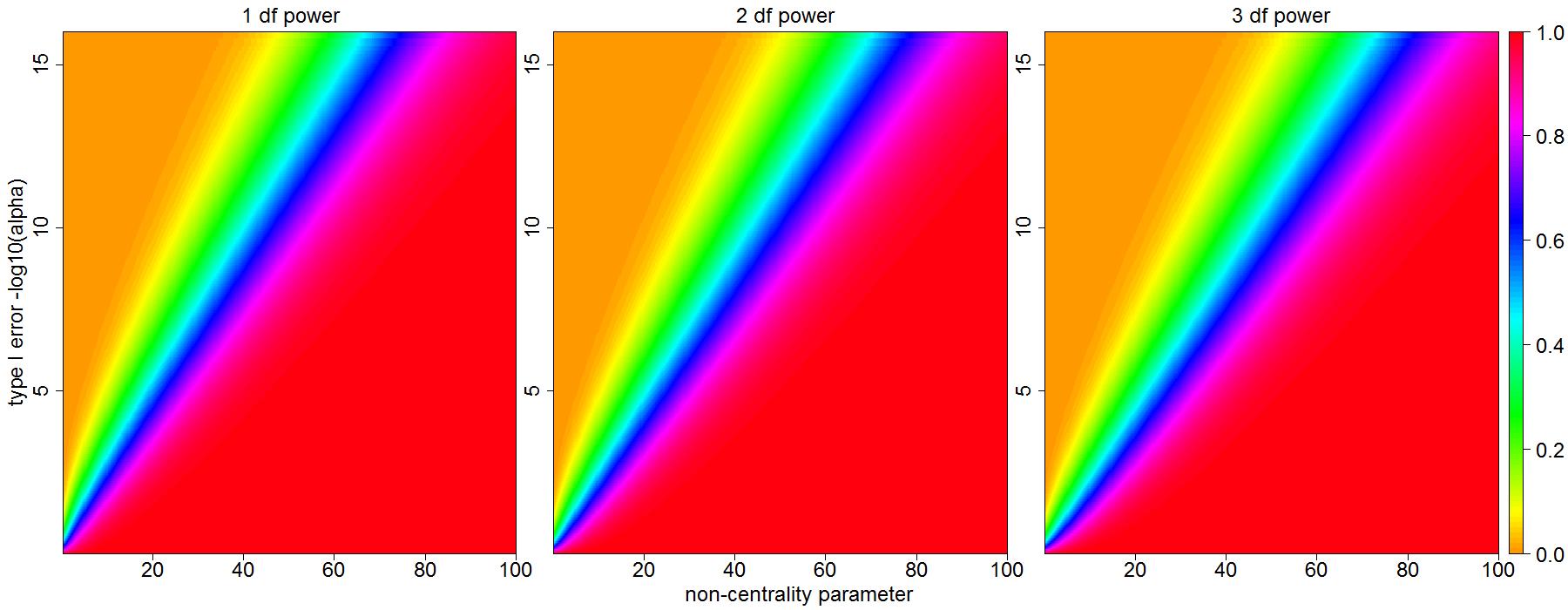}}
 \centerline{\includegraphics[scale=0.25]{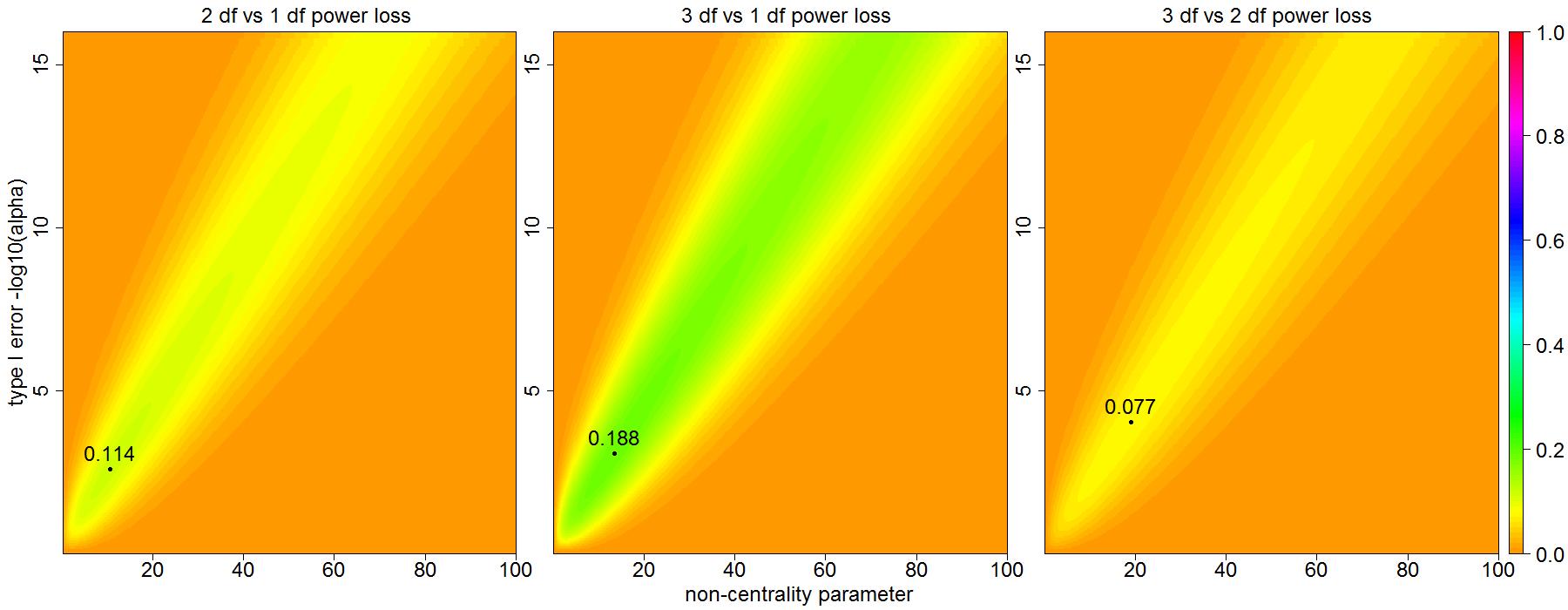}}
\caption{{\bf Heat plots of power and power loss for 1, 2 and 3 degree of freedom chi-squared distributions.} \textbf{Upper panels:} Power of $W_1 \sim \chi^2_{(1,ncp)}$, $W_2 \sim \chi^2_{(2,ncp)}$ and $W_3 \sim \chi^2_{(3,ncp)}$ as a function of $-\log_{10} \alpha$ (type I error) and non-centrality parameter. \textbf{Lower panels:} Power loss of $W_2$ vs $W_1$, $W_3$ vs $W_1$ and $W_3$ vs $W_2$ as a function of $-\log_{10} \alpha$ and non-centrality parameter assuming equal non-centrality parameter within each pair. Black dots correspond to maximum power loss: $\alpha=0.0025$ and $ncp=10.6$ for left panel, $\alpha=0.0008$ and $ncp=13.4$ for middle panel and $\alpha=9.12 \times 10^{-5}$ and $ncp=19$ for right panel.}
\label{heat}  
\end{figure}

Suppose $W_1, W_2$ and $W_3$ follow 1, 2 and 3 degree of freedom chi-squared distribution with common non-centrality parameter $ncp$. Figure \ref{heat} provides heat plots for power and pairwise power loss for $W_1, W_2$ and $W_3$ as functions of $ncp$ and type I error $\alpha$.

As discussed in Section 3.1, power comparison between $W_2$ vs. $W_1$ is more relevant to an autosomal variant. Using the genotype codings $G_A=(0,1,2)$ and $G_D=(0,1,0)$ introduced in Section 1 for an autosomal SNP, we define the additive and genotypic models using the generalized linear regression models \citep{mccullagh89}. Let $g$ be the link function, the additive model is 
\begin{equation}
\label{eq:1}
g(E(Y))=\beta_0+\beta_AG_A,
\end{equation}
and the genotypic model is 
\begin{equation}
\label{eq:2}
g(E(Y))=\beta_0+\beta_AG_A+\beta_DG_D,
\end{equation}
Then $W_1$ is the standard association test statistic for testing $H_0: \beta_A=0$ based on the additive model (\ref{eq:1}), and $W_2$ is the test statistic for testing $H_0: \beta_A=0 \mbox{ and } \beta_D=0$ jointly using the genotypic model (\ref{eq:2}).  Figure \ref{heat} shows the maximum power loss of a 2.d.f genotypic test is 11.4\%.

Notably the maximum of 11.4\% power loss holds for comparing any 2 d.f.\ $\chi^2$ test with a 1 d.f.\ $\chi^2$ test that was derived from the correctly specified 1 d.f.\ model. This is because the derivation is based on $ncp$ and $\alpha$ alone. For instance, for the association analysis of interest here, if the true genetic model is recessive (i.e.\ 1 d.f.) then power loss of using the 2 d.f.\ genotypic test compared to using the true recessive model is also capped at 11.4\%. Second, we emphasize that although a 11.4\% loss of power is substantial, the fact that this is the maximum power loss for the 2 d.f.\ genotypic test, under any 1 d.f. genetic models and regardless of the true genetic effect size, sample size and significance level, is encouraging, as the potential power gain of the genotypic test under other models can be much greater than 11.4\% which we study next. 

\begin{figure}
 \centerline{\includegraphics[scale=0.3]{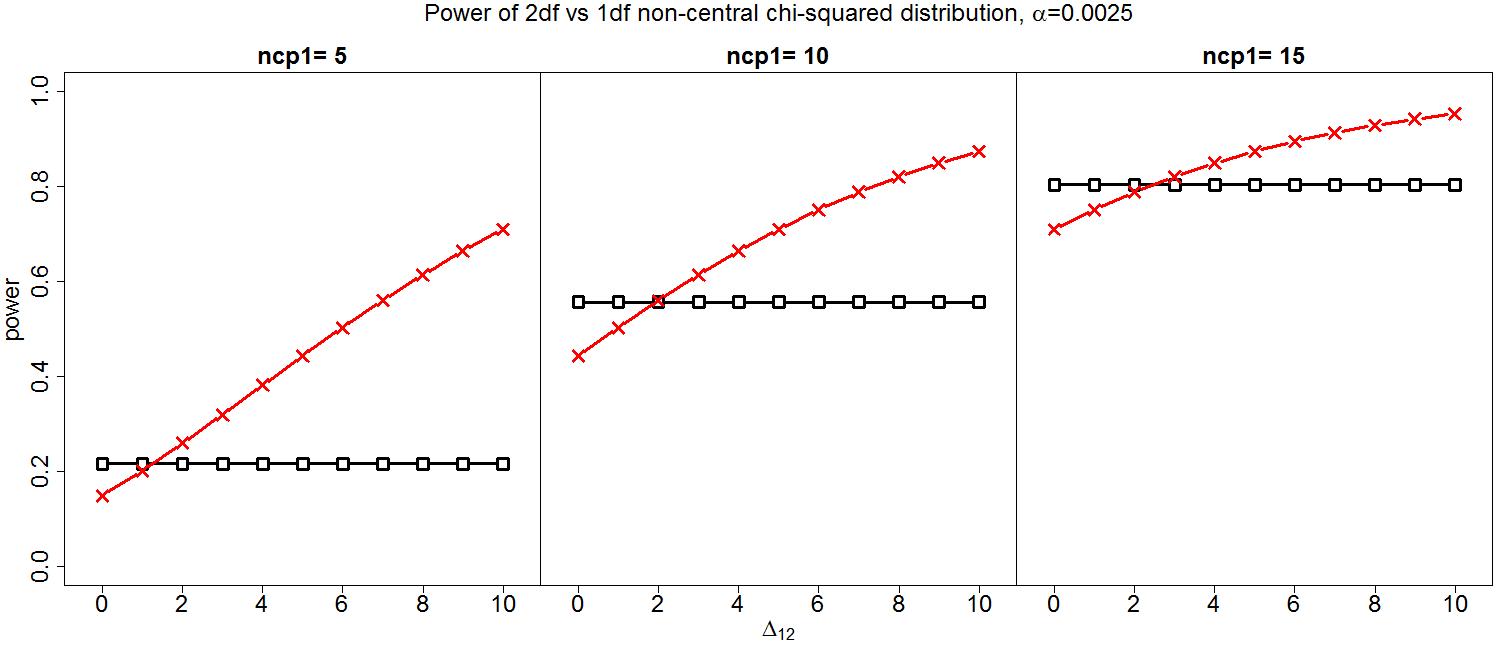}}
 \centerline{\includegraphics[scale=0.3]{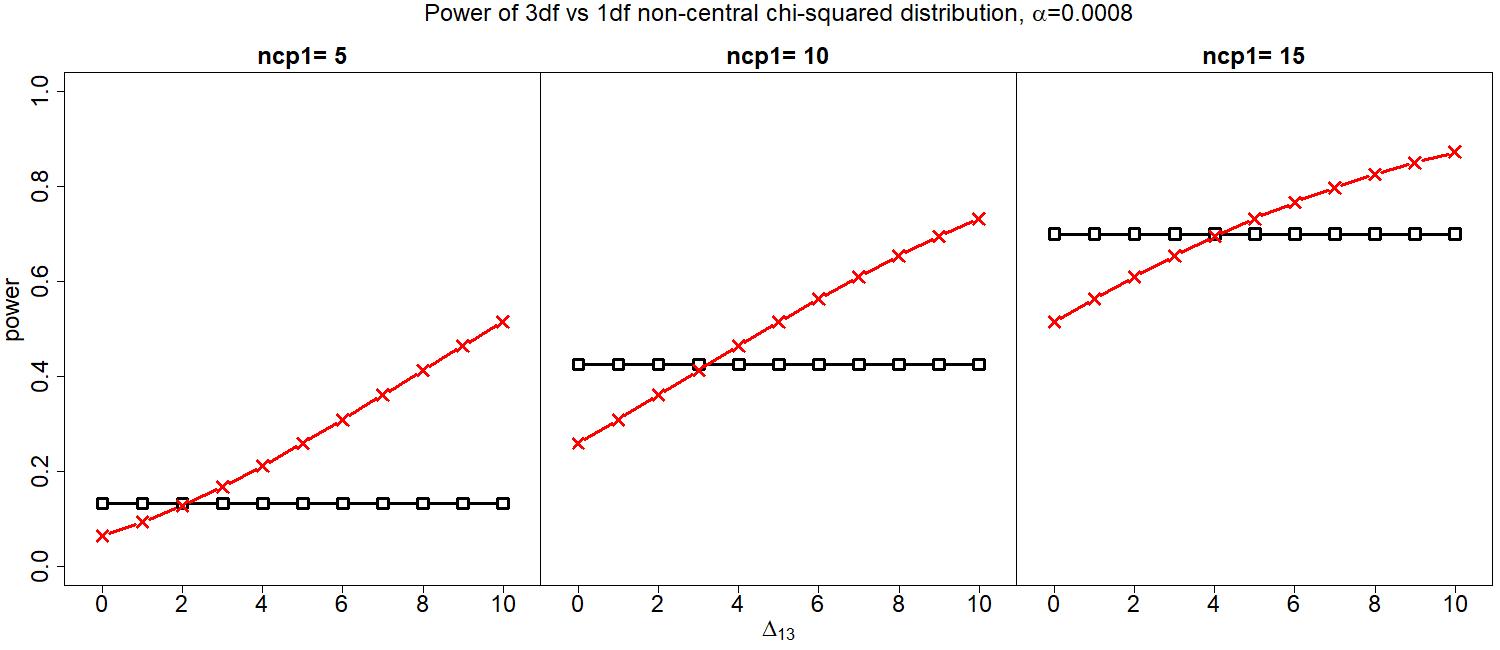}}
\caption { {\bf Power gain vs. power loss compared to $\boldsymbol{W_1 \sim \chi^2_{(1,ncp_1)}}$ when $\boldsymbol{ncp_1=5, 10}$ or 15 under the worst case scenarios (type I error $\boldsymbol{\alpha}$ leading to the maximum power loss). Upper panels:} $W_2 \sim \chi^2_{(2,ncp_2)}$, $\Delta_{12}=ncp_2-ncp_1$ varying from 0 to 10 and $\alpha=0.0025$. Black $\square$ curves are power of $W_1$; {\color{red} red $\times$ curves} are power of $W_2$. 
{\bf Lower panels:} $W_3 \sim \chi^2_{(3,ncp_3)}$, $\Delta_{13}=ncp_3-ncp_1$ varying from 0 to 10 and $\alpha=0.0008$. Black $\square$ curves are power of $W_1$; {\color{red} red $\times$ curves} are power of $W_3$.
}
\label{gain}  
\end{figure}

In Section 3.1, we discussed the power gain vs. power loss between the 3 d.f. test and 1 d.f. test for an X-chromosome SNP based on results in Figure \ref{gain}. Figure \ref{gain} also shows the power comparison between 2 d.f. and 1 d.f. test for autosome. In the presence of a dominance effect, $ncp_2=ncp_1+\Delta_{12}$, where $\Delta_{12}>0$ and the value depends on sample size and the degree of departure from additivity. Compared to the maximum power loss of using the genotypic test under additivity, the maximum power gain under non-additivity can be theoretically as large as $1-\alpha$ (i.e. close to 100\%). To provide specific numerical results, we consider $\alpha=0.0025$ (the worse-case scenario derived above for the 2 d.f.\ test), $ncp_1=5$, 10 or 15, and $\Delta_{12}$ ranging from 0 to 10.  Results show that once $\Delta_{12}$ is as large as a quarter of $ncp_1$ (i.e.\  $ncp_2 \approx 1.25 \cdot ncp_1$), the power gain is bigger than power loss when $\Delta_{12}=0$. This observation together with the 11.4\% maximum power loss suggest that the genotypic model should be considered for use in association studies of autosomal SNPs.

We have shown power computation using the general theory of chi-squared distributions. To compute power under a specific genetic model, we will need to compute $ncp$ based on the model specific regression coefficient $\boldsymbol{\beta}$. Suppose $\boldsymbol{\beta}$ has length $p$. We then partition $\boldsymbol{\beta}=(\beta_1', \beta_2')'$ corresponding to, respectively, the $(p-q)$ secondary covariates not being tested and the $q$ primary covariates of interest. For instance, $\beta_2=(\beta_A, \beta_D)'$ for the genotypic model (\ref{eq:2}) of an autosomal SNP, and $\beta_2=(\beta_A, \beta_D, \beta_{GS})'$ for the 3 d.f. Model $M_4$ of an X-chromosome SNP. To distinguish between the combined regression coefficients in vector form and its partition, we use bold symbols ($\boldsymbol{\beta}$) to denote all parameters in a regression, and non-bold symbols ($\beta$) to denote each partition throughout our work, although each partition may have dimension $> 1$. 
To compute the non-centrality parameter, we first partition the expected Fisher information matrix accordingly,
 $$H(\beta_1, \beta_2)=\left[ \begin{array}{cc}
            H_{11}(\beta_1, \beta_2) & H_{12}(\beta_1, \beta_2)  \\
            H_{21}(\beta_1, \beta_2) & H_{22}(\beta_1, \beta_2) \\
               \end{array} \right].$$
We can then obtain the non-centrality parameter, where
\begin{eqnarray} 
ncp&=&\beta_2'[H_{22}(\beta_1, 0)-H_{21}(\beta_1,0)H_{11}^{-1}(\beta_1, 0)H_{12}(\beta_1, 0)]\beta_2.
\label{ncp_exact}
\end{eqnarray}

The technical details for computing this $ncp$ for different genetic models are given in Appendix C. Note that when a dominance and/or interaction effect is present but the additive model is used, the computation of the non-centrality parameter is less straightforward because of the model misspecification.  Although the derivation is difficult under the standard genotype coding, a re-parametrization can considerably simplify the computation \citep{begg92}. We provide detailed discussion for non-centrality parameter computation under misspecified genetic models in Appendix D.

Figure \ref{Apower} shows power of the two tests (the 1 d.f.\ additive test and 2 d.f.\ genotypic test) for association analyses of an autosomal SNP across a range of dominance effects including no dominance effect, and when $n=1,000$ and $\alpha=0.0025$. Power of the two tests were computed based on analytically derived $ncp$s and further validated by simulation studies, and the corresponding non-centrality parameter values are provided in Figure \ref{Ancp}. We fixed the additive effect at $\beta_A=0.3$ while varying $\beta_D$ from $-0.6$ to $0.6$, covering a wide range of genetic models. For example,
$\beta_D=0$, $\beta_D=0.3$ and $\beta_D=-0.3$ represent, respectively, additive, dominant and recessive models, while $\beta_D>0.3$ and $\beta_D<-0.3$ cover, respectively, over- and under-dominant models.  We considered a wide range of common allele frequencies between 0.1 and 0.9 but present results only for $f=0.2, 0.5$ and 0.8 which are representative of all results; analysis of multiple rare variants jointly is beyond the scope of this work. Table \ref{ncptable} shows additional scenarios of different additive and dominance effect for a more intuitive interpretation of $ncp$ under additive and genotypic model.

\begin{table}
\centering
\begin{tabular}{ccc|cc}
  \hline
$\beta_A$ & $\beta_D$ & $f$ & Additive & Genotypic \\ 
  \hline
0.3 & 0 & 0.2 & 7.2 & 7.2 \\ 
0.6 & 0 & 0.2 & 28.8 & 28.8 \\
0 & 0.3& 0.2 & 2.592 & 4.896 \\
0 & 0.6 & 0.2 & 10.368 & 19.584 \\
0.3 & 0.3 & 0.2 & 18.432 & 20.736 \\
0.3 & -0.3 & 0.2 & 1.152 & 3.456 \\ \hline 
0.3 & 0 & 0.5 & 11.25 & 11.25 \\ 
0.6 & 0 & 0.5 & 45 & 45 \\
0 & 0.3& 0.5 & 0 & 5.625 \\
0 & 0.6 & 0.5 & 0 & 22.5 \\
0.3 & 0.3 & 0.5 & 11.25 & 16.875 \\
0.3 & -0.3 & 0.5 & 11.25 & 16.875 \\ \hline 
\end{tabular}
\caption{{\bf Non-centrality parameter values of additive model and genotypic model under different scenarios.} Note that the value of $ncp$ depends on additive effect $\beta_A$, dominance effect $\beta_D$, allele frequency $f$, and sample size $n$. Here $n$ is fixed at 1000; it is straightforward to calculate $ncp$ for other sample sizes, because $ncp$ is proportional to $n$.}
\label{ncptable}
\end{table}

\begin{figure}
 \centerline{\includegraphics[scale=0.3]{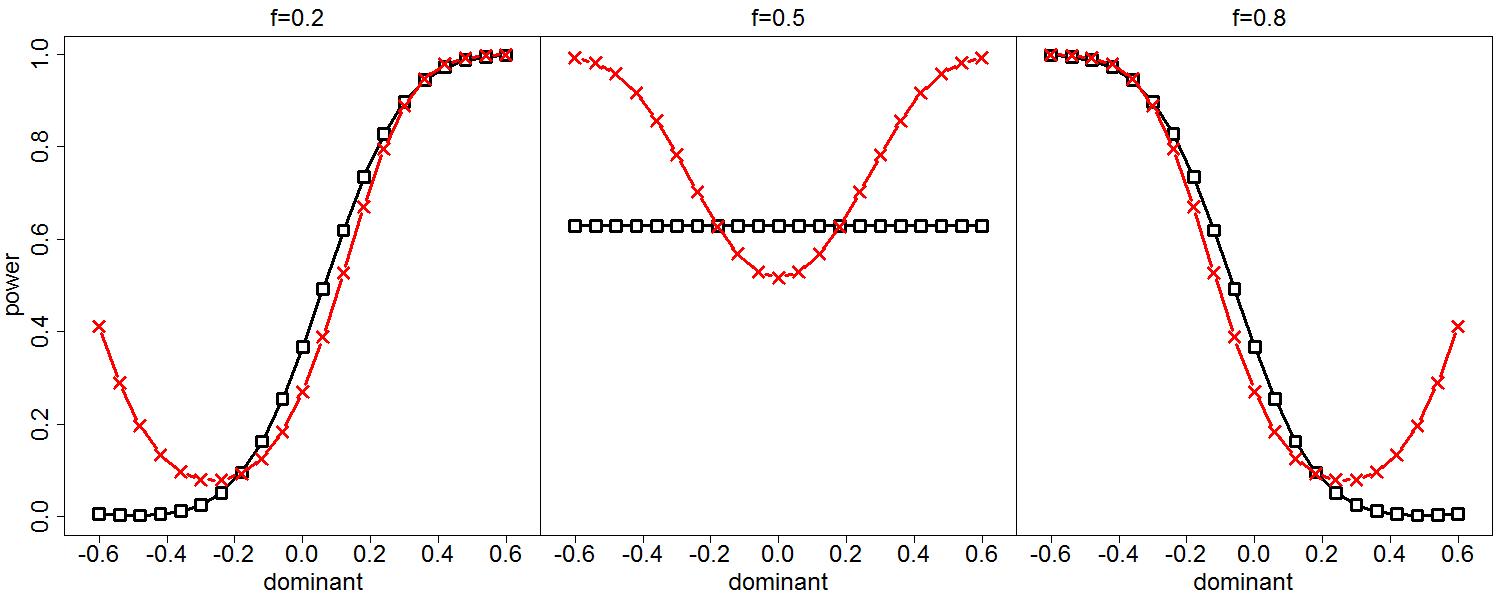}}
\caption{{\bf Power comparison between the additive and genotypic tests for association analyses of autosomal SNPs across a range of dominance effects, including no dominance effect.} The additive effect is fixed at $\beta_A=0.3$, while the dominance effect $\beta_D$ ranges from $-0.6$ to $0.6$.  The allele frequency $f=0.2, 0.5$ and 0.8 for the three plots, respectively, from left to right, the sample size $n=1,000$, and the size of the test $\alpha=0.0025$. The black $\square$ curves are power of testing $\beta_A=0$ using the additive model (\ref{eq:1}),  and the {\color{red} red $\times$ curves} are power of testing $\beta_A=\beta_D=0$ using the genotypic model (\ref{eq:2}).} 
\label{Apower}  
\end{figure}

\begin{figure}
 \centerline{\includegraphics[scale=0.3]{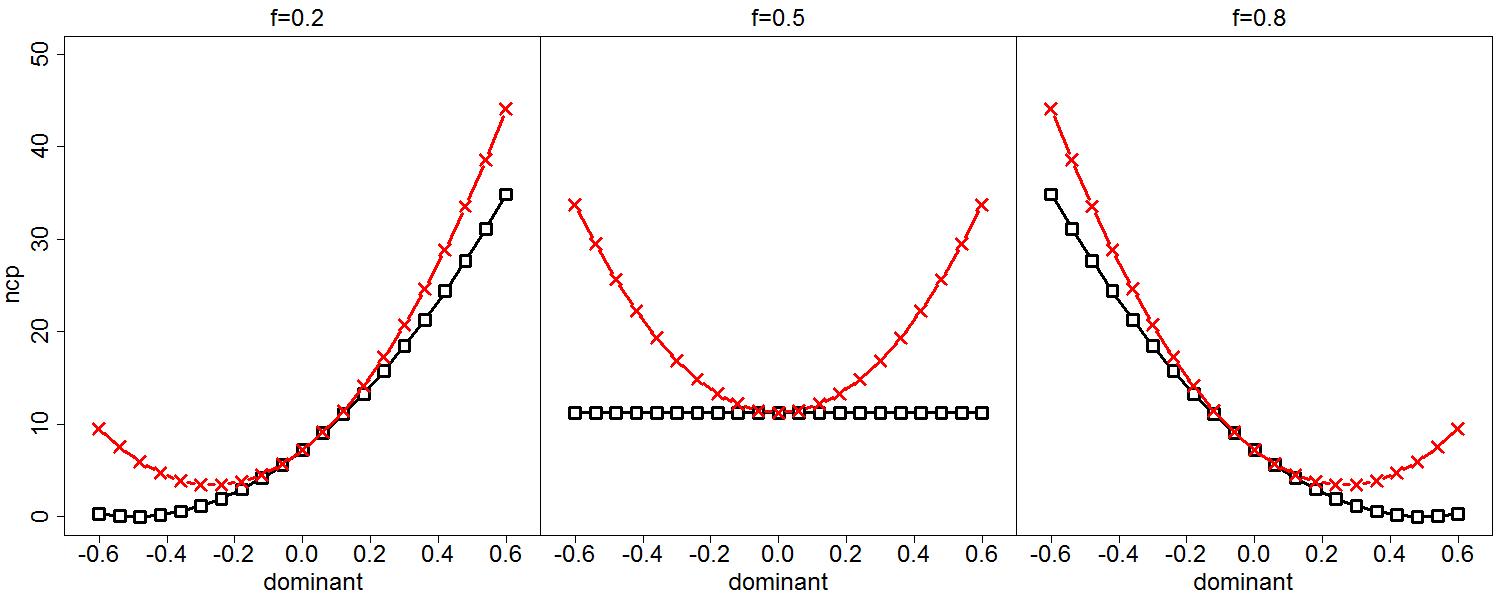}}
\caption{{\bf Non-centrality parameter comparison between the additive and genotypic tests for association analyses of autosomal SNPs across a range of dominant effects, including no dominant effect.}  The additive effect is fixed at $\beta_A=0.3$, while the dominant effect $\beta_D$ ranges from $-0.6$ to $0.6$.  The allele frequency $f=0.2, 0.5,$ and 0.8 for the three plots, respectively, from left to right, the sample size $n=1,000$, and the size of the test $\alpha=0.0025$. The black $\square$ curves are $ncp$ of testing $\beta_A=0$ using the additive model (\ref{eq:1}),  and the {\color{red} red $\times$ curves} are $ncp$ of testing $\beta_A=\beta_D=0$ using the genotypic model (\ref{eq:2}).}
\label{Ancp}  
\end{figure}

The results show that the power gain of using the genotypic test (the red $\times$ curve) as compared with the additive test (the black $\square$ curve) can be more than 40\%; the increase can be bigger for other settings, e.g.\ $\beta_D$ outside the $-0.6$ to $0.6$ range (results not shown). In contrast, we have shown the maximum power loss of using the genotypic test under additivity is never more than 11.4\% analytically.  Indeed, when $f=0.5$ and $\beta_D=0$, $ncp_1=ncp_2=ncp=11.25$ close to 10.6. Recall that $ncp=10.6$ combined with $\alpha=0.0025$ result in maximum power loss of the genotypic test under additivity (Figure \ref{heat}).  Thus, the maximum power loss shown in Figure \ref{Apower} is close to the global maximum possible, but the power gain seen here can be made bigger by considering other scenarios (e.g.\ $\beta_D>0.6$ or $<-0.6$). 

Although the simulated genetic models cover a wide range of genetic models, it is difficult to argue if the maximum power gain for the 2 d.f.\ test is attainable in practice. However, applications in Section 4.2 show that the proposed 2 d.f.\ genotypic test can lead to p-values that are several orders of magnitude smaller than the standard 1 d.f.\ additive test for associated autosomal SNPs. Results here for the autosomes also suggest that investigation of the dominance effect for the X-chromosome may be warranted. 

Figure \ref{Apower} also shows some interesting patterns between power and allele frequencies.  When $f=0.5$ (middle plot), it is straightforward to see why, for the genotypic test (the red $\times$ curve), the minimum power occurs at $\beta_D=0$.  Interestingly, power of the additive test (the black $\square$ curve) in this case is constant across the range of $\beta_D$.  Let $-a$, $d$ and $a$ be the true effect sizes of the three genotypes, $rr$, $rR$ and $RR$, respectively.  The additive effect captured by the additive component is $$a^*=w_{rr} (d-(-a)) + w_{RR} (a-d) = a+ (w_{rr}-w_{RR})\cdot d,$$ where the weighting factors, $w_{rr}$ and $w_{RR}$, are proportional to the corresponding genotype frequencies, $(1-f)^2$ and $f^2$, respectively. When $f=0.5$, $w_{rr}=w_{RR}$ and $a^*\equiv a$ which explains why power of the additive test in this plot is constant across the range of dominant effects.  Similarly,  when $f=0.2$ (left plot), $w_{rr}> w_{RR}$ and $a^*$ increases as $\beta_D$ increases from -0.6 to 0.6, resulting in increased power of the additive test.  Power of the genotypic test also depends on the absolute size of $\beta_D$, which leads to the non-monotone pattern shown in the plot.  Results of $f=0.8$ mirror those of $f=0.2$, as expected. 

For X-chromosome, we provide detailed discussion of power comparisons among the genetic model $M_1-M_4$ in Section 3.2. Results for differential allele $f$ between males and females are shown in Figures \ref{more powerX}. Intuitively explaining the power pattern observed in Figure 2 and Figures \ref{more powerX} for X-chromosome SNPs is more challenging due to the modelling of the interaction effect, but the idea is similar. For example, Figure 2B shows that when $f=0.5$ and interaction effect is fixed at -0.6, $M_3$ (jointly testing additive and interaction effects; the blue $\triangledown$ curve) has constant power across different dominant effects. This is because $\beta_D$ is only relevant to females, so we can use the argument above for the autosomes here. In contrast,$M_2$ jointly tests additive and dominant effects (the green $\diamond$ curve), so its power increases as $\beta_D$ moves away from zero. 

Similarly, when $f=0.2$ (Figure 2A) and interaction effect is fixed at -0.6, $M_3$ (jointly testing additive and interaction effects; the blue $\triangledown$ curve) does not have constant test power. This is because when $f=0.2$ one could consider the additive effect in females as the difference between $\mu_{rr}$ and $\mu_{rR}$; genotype $RR$ is rare with 4\% frequency. On the other hand, varying $\beta_D$ is equivalent to varying $\mu_{rR}$ as noted in Section 3.2, thus power of $M_3$ increases as $\beta_D$ increases even though the interaction effect is fixed at -0.6. This result also highlight the fact that GWAS alone cannot be used to determine the underlying true genetic model.

\section*{Appendix C: Non-centrality Parameter Computation for Correctly Specified Genetic Models}

We provide the details for computing non-centrality parameters for the tests under different genetic models. When the model is correctly specified, $ncp$ may be computed using Equation \ref{ncp_exact} as described in Appendix B. Equation \ref{ncp_exact} computes exact $ncp$ as a function of design matrix $X$. In order to disentangle Equation \ref{ncp_exact} from the sample-specific observed genotypes, we consider the asymptotic behaviour of $ncp$ as $n \to \infty$.  In order to avoid the uninteresting case in which $ncp \rightarrow \infty$ when $n$ grows, we assume $\boldsymbol{\beta}=\boldsymbol{c}/\sqrt{n}$ \citep[see also][]{cox74,begg92,begg93,neuhaus98}  for a fixed vector $\boldsymbol{c}$, so that $\boldsymbol{\beta} \to \mathbf{0}$ and $ncp$ converges to a finite number as $n \to \infty$.

In the case of a linear model with covariate matrix $X$, $H=\frac{X'X}{\sigma^2}$ regardless of $\boldsymbol{\beta}$. Let $P$ be the limit of $\frac{X'X}{n}$:
\begin{eqnarray*}
\frac{X'X}{n} &\overset{p}{\to} P.
\end{eqnarray*}

Corresponding to the split  $\boldsymbol{\beta}=(\beta_1,\beta_2)$, $P$ is partitioned as $P=\left[ \begin{array}{cc}
            P_{11} & P_{12}  \\
            P_{21} & P_{22} \\
               \end{array} \right]$.
The asymptotic value of $ncp$ is then computed following Equation \ref{ncp_exact}:
$$ncp_{(linear)} \overset{p}{\to} \frac{1}{\sigma^2}c_{2}'[P_{22}-P_{21}(P_{11})^{-1}P_{12}]c_{2},$$
where $c_2=\beta_2\sqrt{n}$.

In the logistic model, $H(\boldsymbol{\beta})=X'W(\boldsymbol{\beta})X$, where $W(\boldsymbol{\beta})$ is the $n \times n$ diagonal matrix with the $ith$ diagonal element equal to $\mu_i(\boldsymbol{\beta})(1-\mu_i(\boldsymbol{\beta}))$, and $\mu(\boldsymbol{\beta})=\frac{\exp(X\boldsymbol{\beta})}{1+\exp(X\boldsymbol{\beta})}$. As $n \to \infty$, $\boldsymbol{\beta} \to \mathbf{0}$ so that $\mu_i(\boldsymbol{\beta})(1-\mu_i(\boldsymbol{\beta})) \overset{p}{\to} \frac{1}{4}$, which implies
$$\frac{X'W(\boldsymbol{\beta})X}{n} \overset{p}{\to} \frac{P}{4}.$$
Hence, the asymptotic non-centrality parameter under logistic model is
$$ncp_{(logistic)} \overset{p}{\to} \frac{1}{4}c_{2}'[P_{22}-P_{21}(P_{11})^{-1}P_{12}]c_{2}.$$
Note that if $\sigma^2=4$, the linear and logistic model have equal asymptotic $ncp$ as long as  $X$ and $\boldsymbol{\beta}$ are the same. Under this scenario, in both models
\begin{eqnarray}
ncp \overset{p}{\to} \frac{1}{\sigma^2}c_{2}'[P_{22}-P_{21}(P_{11})^{-1}P_{12}]c_{2}. 
\label{ncp_asymptotic}
\end{eqnarray}
This observation allows a  convenient derivation of  $ncp$ in logistic models by plugging $\sigma^2=4$ in the $ncp$ formula for linear models.

{\it Remark 1}: Assume that  the generative model is genotypic (for autosomal SNPs) or model $M_4$ in Table 3 (for X-chromosome SNPs). If the additive model  or one of the models $M_1, M_2, M_3$ are used for estimation, then the above derivation of $ncp$ is not valid since the estimators for  $\boldsymbol{\beta}$  may be biased due to model misspecification.   

However, when the true model is additive or one of models $M_1, M_2, M_3$, the derivation for $ncp$ remains valid when using either the genotypic model or $M_4$ for estimation, as the MLE estimators of $\boldsymbol{\beta}$ remain unbiased. Therefore, $ncp$ under the genotypic or $M_4$ model may be computed by Equation \ref{ncp_exact} or \ref{ncp_asymptotic} regardless of the true model. 

{\it Remark 2}: Under genotypic model, $\beta_2=(\beta_A, \beta_D)'$, and 
\begin{eqnarray*}
P &=&\left( \begin{array}{ccc}
            1 & E(G_A) & E(G_D) \\
            E(G_A) & E(G_A^2) & E(G_A \cdot G_D) \\
            E(G_D) & E(G_A \cdot G_D) &E(G_D^2) \\
               \end{array} \right).
\end{eqnarray*}
Under $M_4$, $\beta_2=(\beta_A, \beta_D, \beta_{GS})'$ and 
\begin{eqnarray*}
P &=&\left( \begin{array}{ccccc}
            1 & E(S) & E(G_A) & E(G_D) & E(GS) \\
            E(S) & E(S^2) & E(S \cdot G_A) & E(S\cdot G_D) & E(S \cdot GS) \\
            E(G_A) & E(S \cdot G_A) &E(G_A^2) & E(G_A \cdot G_D) & E(G_A \cdot GS) \\
            E(G_D) & E(S \cdot G_D) & E(G_A \cdot G_D) & E(G_D^2) & E(G_D \cdot GS) \\
            E(GS) & E(S \cdot GS) & E(G_A \cdot GS) & E(G_D \cdot GS) & E(GS^2) \\
               \end{array} \right). 
\end{eqnarray*}
Assuming equal population frequency of females and males, $E(S)=0.5$. Other expected values are computed from the non-baseline allele frequencies ($f_{female}$ and $f_{male}$). Although different codings of $G_A$ and $G_I$ may lead to different expected values, the test statistics are common (following Theorem 1) thus implying  that the $ncp$ form is asymptotically coding-invariant. 

\section*{Appendix D: Non-centrality Parameter Computation for Misspecified Genetic models}

Under model misspecification, the derivations in Appendix C may not be applicable. In this section, we provide an alternative approach for deriving $ncp$ by reparametrizing the covariates  without changing the test statistics. 
The approach is illustrated with a series of examples. \\
{\bf Example 1: Additive model is misspecified when dominant effect is present}\\
 The following four steps are used to compute the correct $ncp$:
\begin{itemize}
\item[{\bf S1}] We reparametrize $G_A$ and $G_D$ as $G_A^*$ and $G_D^*$ such that the test statistic for the null $H_0: \beta_A^*=\beta_D^*=0$ is the  same  as  that for $H_0: \beta_A=\beta_D=0$ under the original genotypic model. From Theorem 1 it is sufficient that  $(1, G_A^*, G_D^*)$ is a linear transformation of $(1, G_A, G_D)$.

\item[{\bf S2}] We next test $\beta_A^*=0$ under the reparametrized genotypic model $Y \sim G_A^*+G_D^*$. Because the reparametrized genotypic model is correctly specified, the asymptotic $ncp$ for this test can be computed following Equation \ref{ncp_asymptotic}.

\item[{\bf S3}] We show that when $\mbox{corr}(G_A^*, G_D^*)=0$, the re-parametrized additive model: $Y \sim G_A^*$ and genotypic model: $Y \sim G_A^*+G_D^*$ have asymptotic equal $ncp$ for testing $\beta_A^*=0$.

\item[{\bf S4}] We require $(1, G_A^*)$ to be a linear transformation of $(1, G_A)$. Then by Theorem 1, testing $\beta_A^*=0$ under $Y \sim G_A^*$ has the same test statistic as testing $\beta_A=0$ under $Y \sim G_A$. Therefore, the correct $ncp$ for testing $\beta_A=0$ under original additive model $Y \sim G_A$ is asymptotically equal to the $ncp$ computed in step 1.
\end{itemize} 

{\bf S1:} We define $G_A^*=(-1, 0, 1)$ and $G_D^*=(-2f^2, 2f(1-f), -2(1-f)^2)$ for genotype $rr$, $rR$ and $RR$. Direct verification shows $\mbox{corr}(G_A^*, G_D^*)=0$. Also note that $(1, G_A^*, G_D^*)$ and $(1, G_A^*)$ are linear transformations of $(1, G_A, G_D)$ and $(1, G_A)$. Under the new codings, $\boldsymbol{\beta}$ is also re-parametrized so that $\beta_A^*=\beta_A+\beta_D(1-2f)$ and $\beta_D^*=\beta_D$. \\
{\it Remark 3}: Note that the new codings are hard to interpret and we do not recommend using them for effect estimates. Their sole  purpose is to facilitate the asymptotic calculation of  $ncp$.

{\bf S2} See previous section.

{\bf S3} For logistic models, \citet{begg92} showed the equivalence and we apply their conclusion directly. Here we provide  the proof  for the linear model.

Because the re-parametrized genotypic model $Y \sim G_A^*+G_D^*$ is correctly specified, the asymptotic $ncp$ for testing $\beta_A^*=0$ can be computed following Equation \ref{ncp_asymptotic}. With the new coding of $G_A^*$ and $G_D^*$, we have
$$P^*= \left( \begin{array}{ccc}
            1 & -1+2f & 0 \\
            -1+2f & 1-2f+2f^2 & 0 \\
            0 & 0 & 4f^2(1-f)^2 \\
               \end{array} \right).$$
For testing $\beta_A^*=0$, $\boldsymbol{\beta^*}$ is partitioned as $\beta_1^*=(\beta_0^*, \beta_G^*)$ and $\beta_2^*=\beta_A^*$. $P^*$ is partitioned accordingly so that 
$$P_{11}^*= \left( \begin{array}{cc}
            1 & 0 \\
            0 & 4f^2(1-f)^2 \\
               \end{array} \right), P_{21}^*={P_{12}^*}'=\left( \begin{array}{cc}
            -1+2f & 0 \\
               \end{array} \right), P_{22}^*=1-2f+2f^2.$$
Therefore,
$$ncp \overset{p}{\to} \frac{1}{\sigma^2}{c_{2}^*}'[P_{22}^*-P_{21}^*(P_{11}^*)^{-1}P_{12}^*]c_{2}^*=2f(1-f)\frac{n{\beta_A^*}^2}{\sigma^2}.$$

To compute the $ncp$ from the re-parametrized additive model $Y \sim G_A^*$, the model is misspecified so that we need to work on the $ncp$ directly. The chi-squared statistic is 
$$W=\frac{\boldsymbol{\hat{\beta_A^*}}'L'(L({X_A^*}'X_A^*)^{-1}L')^{-1}L\boldsymbol{\hat{\beta_A^*}}}{\sigma^2},$$
where $\boldsymbol{\hat{\beta_A^*}}=(\hat{\beta_0^*}, \hat{\beta_A^*})'$ is the least square estimator of $(\beta_0, \beta_A)'$, $X_A^*=(1, G_A^*)$ and 
$L=\left[\begin{array}{cc}
                   0 & 1  \\
           \end{array}\right]$.

Because the genotypic model is the true model, $Y \sim N(X^*\boldsymbol{\beta^*}, \sigma^2I_n)$, where $X^*=(1, G_A^*, G_D^*)$ and $\boldsymbol{\beta^*}=(\beta_0^*, \beta_A^*, \beta_D^*)'$. It implies that
$$\boldsymbol{\hat{\beta_A^*}}=({X_A^*}'X_A^*)^{-1}{X_A^*}'Y \sim N(({X_A^*}'X_A^*)^{-1}{X_A^*}'X^*\boldsymbol{\beta^*}, ({X_A^*}'X_A^*)^{-1}\sigma^2),$$ 
and thus
$$L\boldsymbol{\hat{\beta_A^*}}\sim N(L({X_A^*}'X_A^*)^{-1}{X_A^*}'X^*\boldsymbol{\beta^*}, L({X_A^*}'X_A^*)^{-1}L'\sigma^2).$$ 
Therefore, $W \sim \chi^2_{(1, ncp_A)}$ where
$$ncp_A=\frac{1}{\sigma^2}\boldsymbol{\beta^*}'{X^*}'X_A^*({X_A^*}'X_A^*)^{-1}L'(L({X_A^*}'X_A^*)^{-1}L')^{-1}
L({X_A^*}'X_A^*)^{-1}{X_A^*}'X^*\boldsymbol{\beta^*}.$$
Next, $X^*\boldsymbol{\beta^*}=X_A^*(\beta_0^*, \beta_A^*)'+\beta_D^*G_D^*$, so we may decompose $ncp_A$ into three parts such that $ncp_A=a_1+a_2+a_3$, where
\begin{eqnarray*}
a_1&=&\frac{1}{\sigma^2}(\beta_0^*, \beta_A^*)L'(L({X_A^*}'X_A^*)^{-1}L')^{-1}L(\beta_0^*, \beta_A^*)', \\
a_2&=&\frac{2}{\sigma^2}\beta_D^*{G_D^*}'X_A^*({X_A^*}'X_A^*)^{-1}L'(L({X_A^*}'X_A^*)^{-1}L')^{-1}L
(\beta_0^*, \beta_A^*)', \\
a_3&=&\frac{1}{\sigma^2}\beta_D^*{G_D^*}'X_A^*({X_A^*}'X_A^*)^{-1}L'(L({X_A^*}'X_A^*)^{-1}L')^{-1}L
({X_A^*}'X_A^*)^{-1}{X_A^*}'G_D^*\beta_D^*.
\end{eqnarray*}
Because 
$$\frac{1}{n}{G_D^*}'X_A^* \overset{p}{\to} (E[G_D^*], E[G_A^*G_D^*])=(0,0)$$ 
and 
$$\boldsymbol{\beta^*}=\frac{\boldsymbol{c^*}}{\sqrt{n}} \overset{p}{\to} \mathbf{0},$$
we have
$a_2 \overset{p}{\to} 0$ and $a_3 \overset{p}{\to} 0$.
To compute $a_1$,
\begin{equation*}
\frac{1}{n}{X_A^*}'X_A^*
=\frac{1}{n}\left( \begin{array}{cc}
            n & \sum G_A^*  \\
            \sum G_A^* & \sum {G_A^*}^2 \\
               \end{array} \right)
\overset{p}{\to} \left( \begin{array}{cc}
            1 & -1+2f \\
            -1+2f & 1-2f+2f^2 \\
               \end{array} \right),
\end{equation*}
which implies
$$nL(X_A'X_A)^{-1}L' \overset{p}{\to} \frac{1}{2f(1-f)}.$$
Therefore,
$$ncp_A \overset{p}{\to} a_1 \overset{p}{\to} 2f(1-f)\frac{n{\beta_A^*}^2}{\sigma^2},$$
which completes the proof that the two asymptotic non-centrality parameters are equal.

{\bf Example 2: $\mathbf{M_1}$, $\mathbf{M_2}$ and $\mathbf{M_3}$ are misspecified models when $\mathbf{M_4}$ is the true model}. 

As in the previous example, the reparametrized coding $(1, S^*, G_A^*, G_D^*, GS^*)$ must be a linear transformation of $(1, S, G_A, G_D, GS)$, and $(1, S^*)$ must also be a linear transformation of $(1, S)$. 

The way to code $G_A$ and $GS$ is not an issue because we can show the equivalency of 
$(1, S, G_A, G_D, GS)$ under each way of coding by applying Theorem 1, as summarized in Figure \ref{Ancp}. The requirements in {\bf S3} and {\bf S4} are discussed below for models $M_1-M_3$.

\begin{itemize}
\item[$\mathbf{M_1}$]  To compute $ncp_1$ for testing $\beta_A=0$ under $M_1$, we require 
$$\mbox{corr}(G_D^*, G_A^*)=\mbox{corr}(GS^*, G_A^*)=0,$$
and $(1, S^*, G_A^*)$ is linear transformation of $(1, S, G_A)$.

\item[$\mathbf{M_2}$]   To compute $ncp_2$ for testing $\beta_A=\beta_D=0$ under $M_2$, we require 
$$\mbox{corr}(GS^*, G_A^*)=\mbox{corr}(GS^*, G_D^*)=0,$$ 
and $(1, S^*, G_A^*, G_D^*)$ is linear transformation of $(1, S, G_A, G_D)$.

 \item[$\mathbf{M_3}$]  To compute $ncp_3$ for testing $\beta_A=\beta_{GS}=0$ under $M_3$, we require 
$$\mbox{corr}(G_D^*, G_A^*)=\mbox{corr}(G_D^*, GS^*)=0,$$ 
and $(1, S^*, G_A^*, GS^*)$ is linear transformation of $(1, S, G_A, GS)$.
\end{itemize}

We can show $ncp_1$, $ncp_2$ and $ncp_3$ are asymptotically equal to the $ncp$s for testing $\beta_A^*=0$, $\beta_A^*=\beta_D^*=0$ and $\beta_A^*=\beta_{GS}^*=0$ under the correctly specified re-parametrized model $M_4$: $Y \sim S^*+G_A^*+G_D^*+GS^*$, which can be computed using Equation \ref{ncp_asymptotic}. The proof under logistic model is a direct application of \citet{begg92}'s result. The proof under linear models is omitted because it is similar to the {\bf Example 1} above but much more lengthy.

The remaining question is to find the re-parametrized codings satisfying the above conditions. We provide such codings in Table \ref{coding2}.

\begin{table}[h]
\begin{center}
\caption{Re-parametrized codings of additive, dominant, interaction and sex effect}
\vspace{2mm}
\label{coding2}
\begin{tabular}{c|c|c|c|c|c} \hline \hline
 & \multicolumn{3}{c|}{Female}&\multicolumn{2}{c}{Male}\\ \cline{2-6}
 Coding & $rr$ & $rR$ & $RR$ & $r$ & $R$ \\ \hline
 $G_A^*$ & -1 & 0 & 1 & -1 & 1 \\ 
 $G_D^*$ & $-2f_{female}^2$ & $2f_{female}(1-f_{female})$ & $-2(1-f_{female})^2$ & 0 & 0 \\
 $GS^*$ & $-f_{female}$ & $\frac{1}{2}-f_{female}$ & $1-f_{female}$ & $\frac{f_{female}(1-f_{female})}{2(1-f_{male})}$ & $-\frac{f_{female}(1-f_{female})}{2f_{male}}$ \\ 
 $S^*$ & -1 & -1 & -1 & 1 & 1 \\ \hline \hline 
\end{tabular}
\end{center}
\end{table} 

{\it Remark 4}: If the missing covariates in the misspecified model are uncorrelated to the covariate being tested, for finite sample it is well-known that the estimator from misspecified model should be unbiased but less efficient. However, we find the $ncp$s are asymptotically equal under the true model and under the misspecified model using re-parametrized codings. This  suggests that the misspecified model is asymptotically as efficient as the true model, in contradiction with the finite sample result. This tension appears because we assume that the nuisance parameter $\beta_1$ also converges to 0. For instance, \citet{begg92} considered the situation in which only $\beta_2$ converges to 0, and their derivations showed that  the asymptotic relative efficiency of the misspecified model and true model is less than 1, in agreement with small sample results.

{\it Remark 5}: The $ncp$ computation in our paper focuses on linear and logistic regressions, but it is possible to extend to other generalized linear models (GLMs). Although there is no result to be directly used for GLMs in general, \citet{neuhaus98} extended \citet{begg92}'s relative efficiency calculation to GLMs. It can be  used to extend the asymptotic equivalence of $ncp$ to other types of GLM.

\section*{Appendix E: Additional Figures}

\begin{figure}
\centering \textbf{A.} $\mathbf{f_{female}=0.2, f_{male}=0.5}$
\begin{center}\includegraphics[scale=0.3]{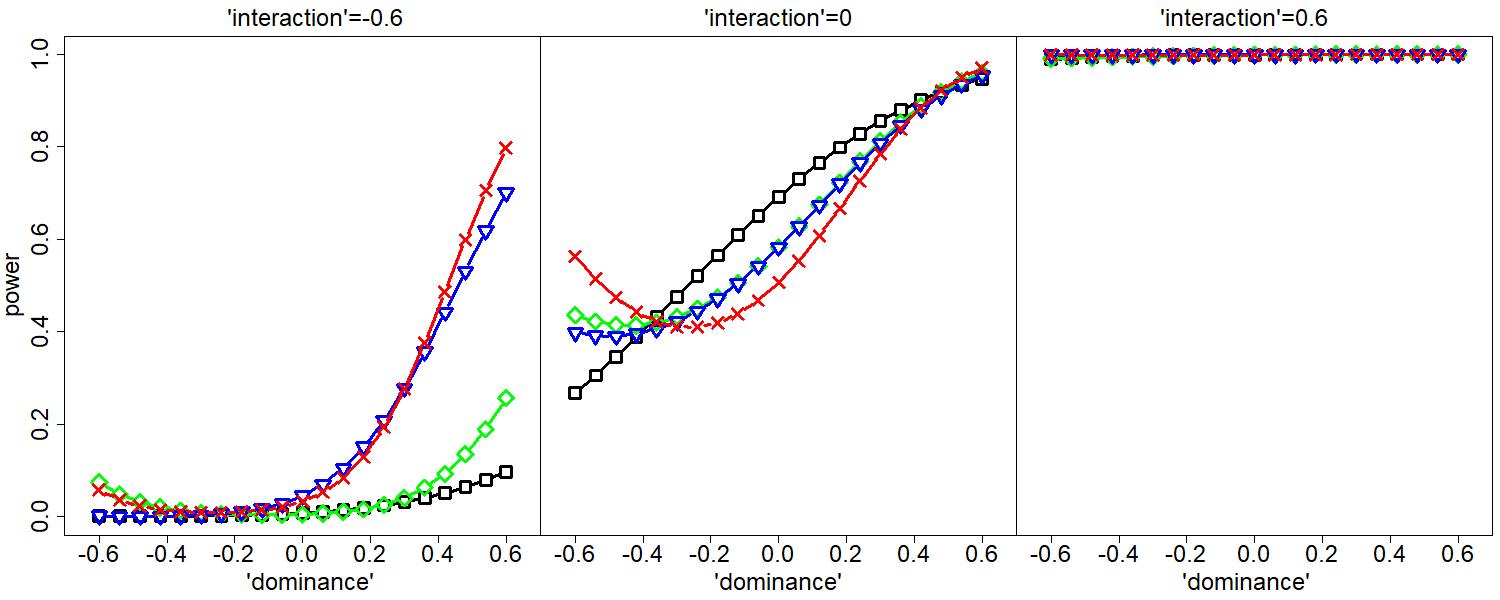}\end{center}
\begin{center}\includegraphics[scale=0.3]{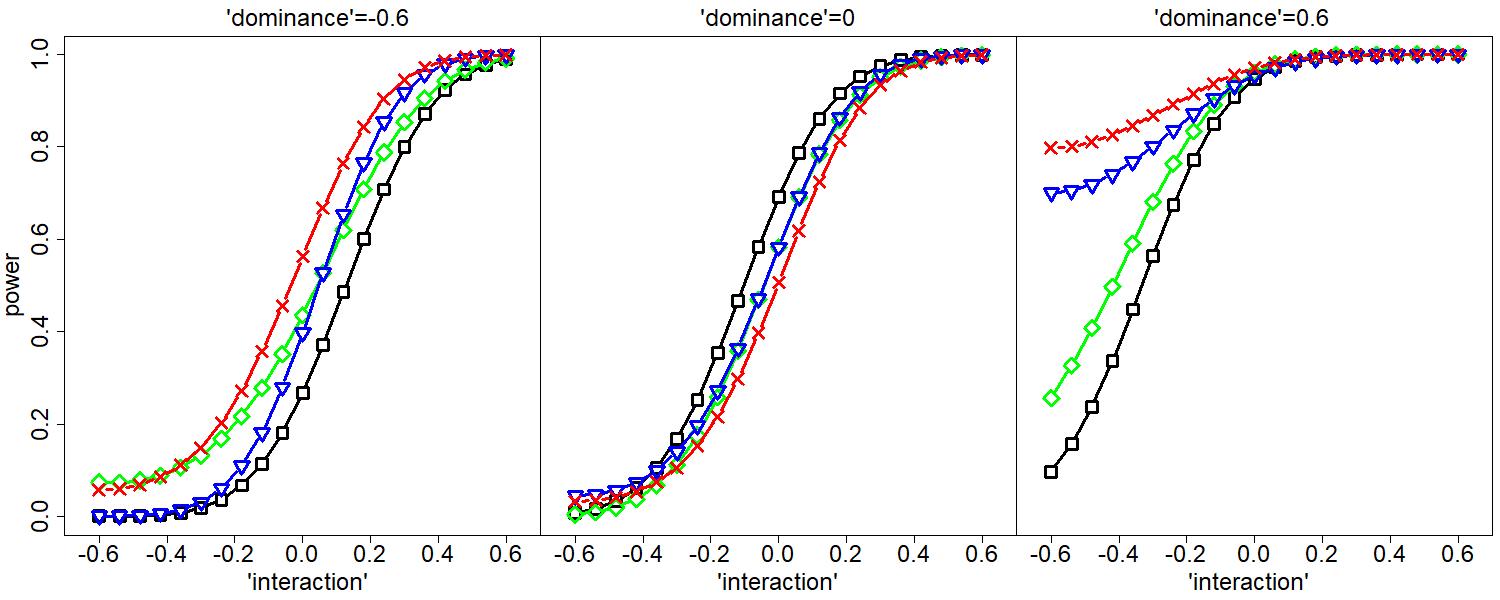}\end{center}
\end{figure}

\begin{figure}
\centering \textbf{B.} $\mathbf{f_{female}=0.2, f_{male}=0.8}$
\begin{center}\includegraphics[scale=0.3]{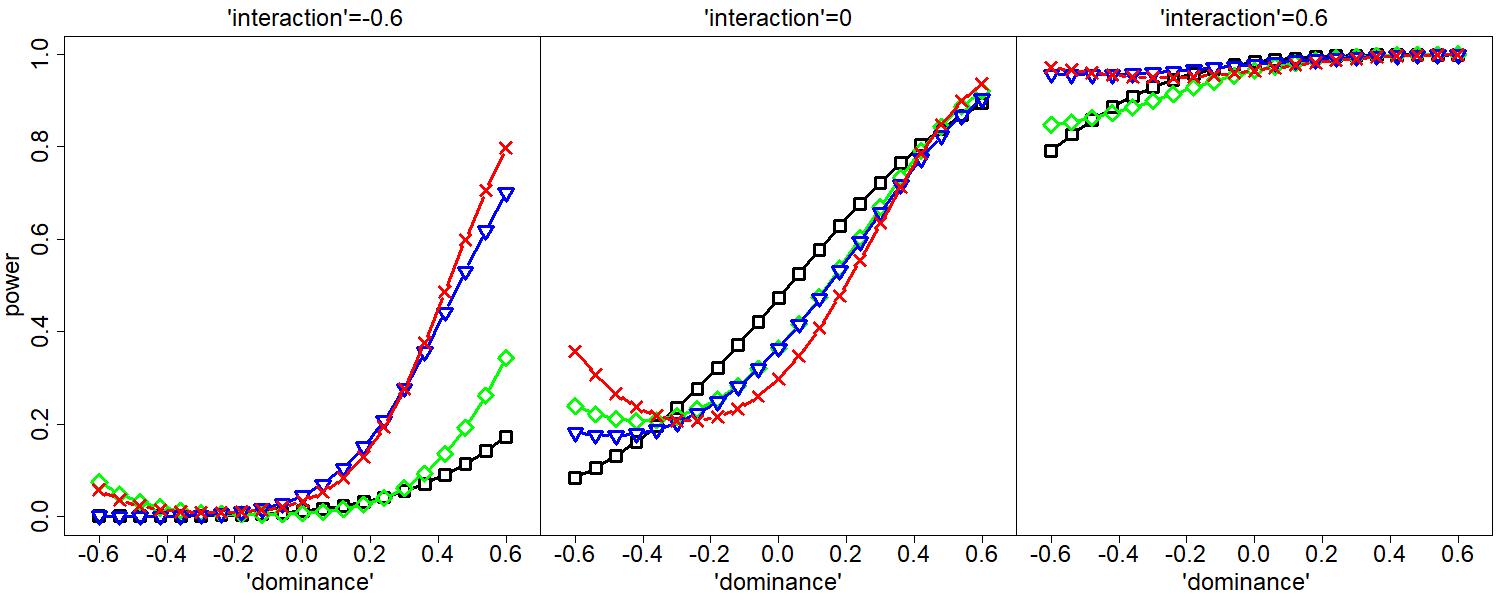}\end{center}
\begin{center}\includegraphics[scale=0.3]{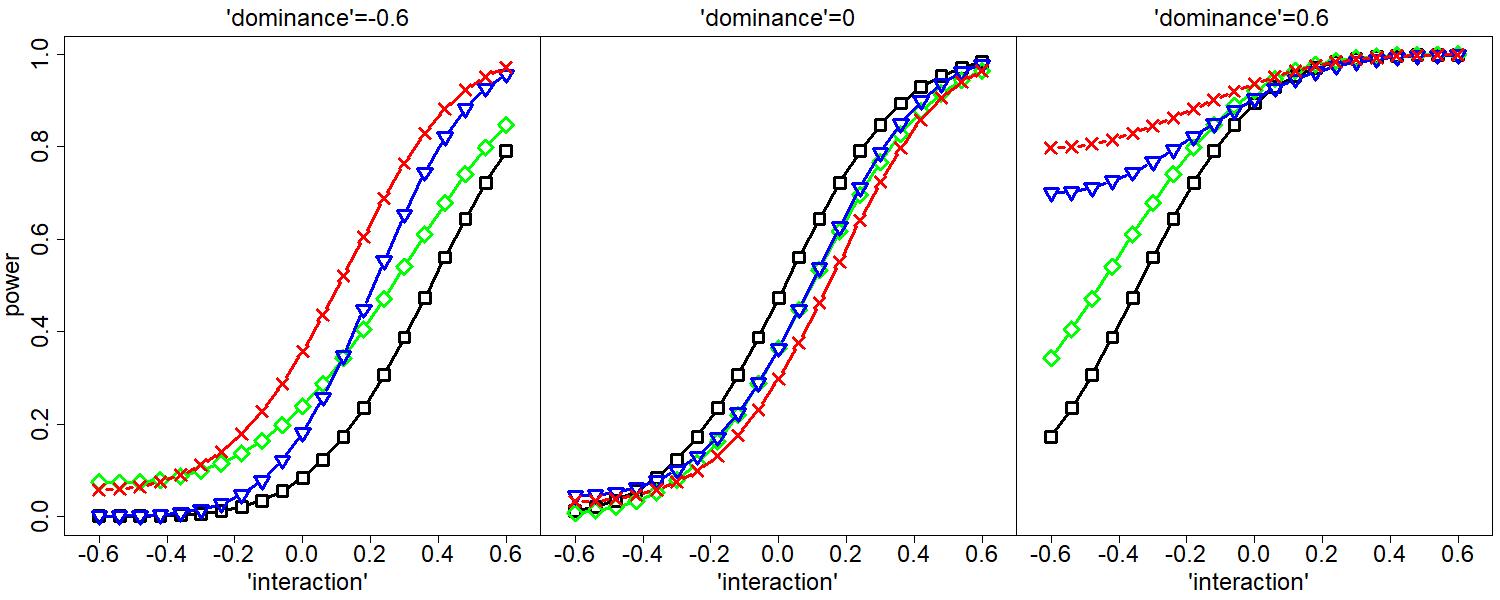}\end{center}
\end{figure}

\begin{figure}
\centering \textbf{C.} $\mathbf{f_{female}=0.5, f_{male}=0.2}$
\begin{center}\includegraphics[scale=0.3]{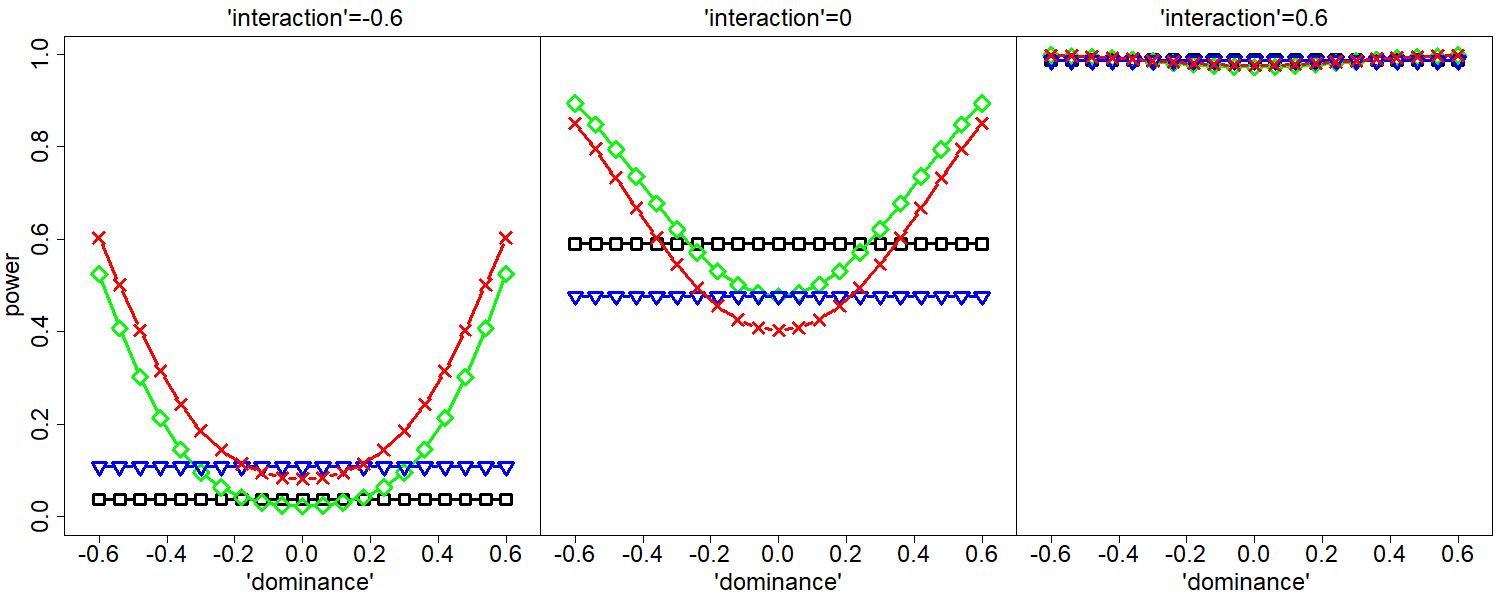}\end{center}
\begin{center}\includegraphics[scale=0.3]{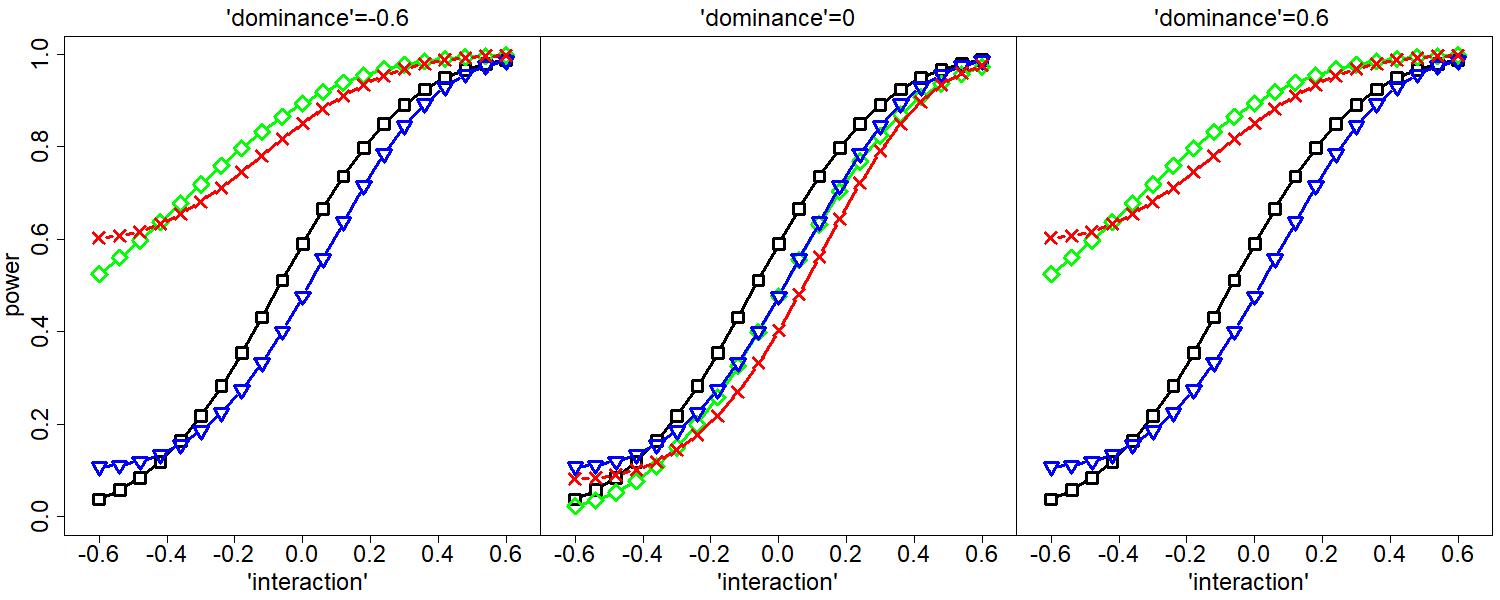}\end{center}
\end{figure}

\begin{figure}
\centering \textbf{D.} $\mathbf{f_{female}=0.5, f_{male}=0.8}$
\begin{center}\includegraphics[scale=0.3]{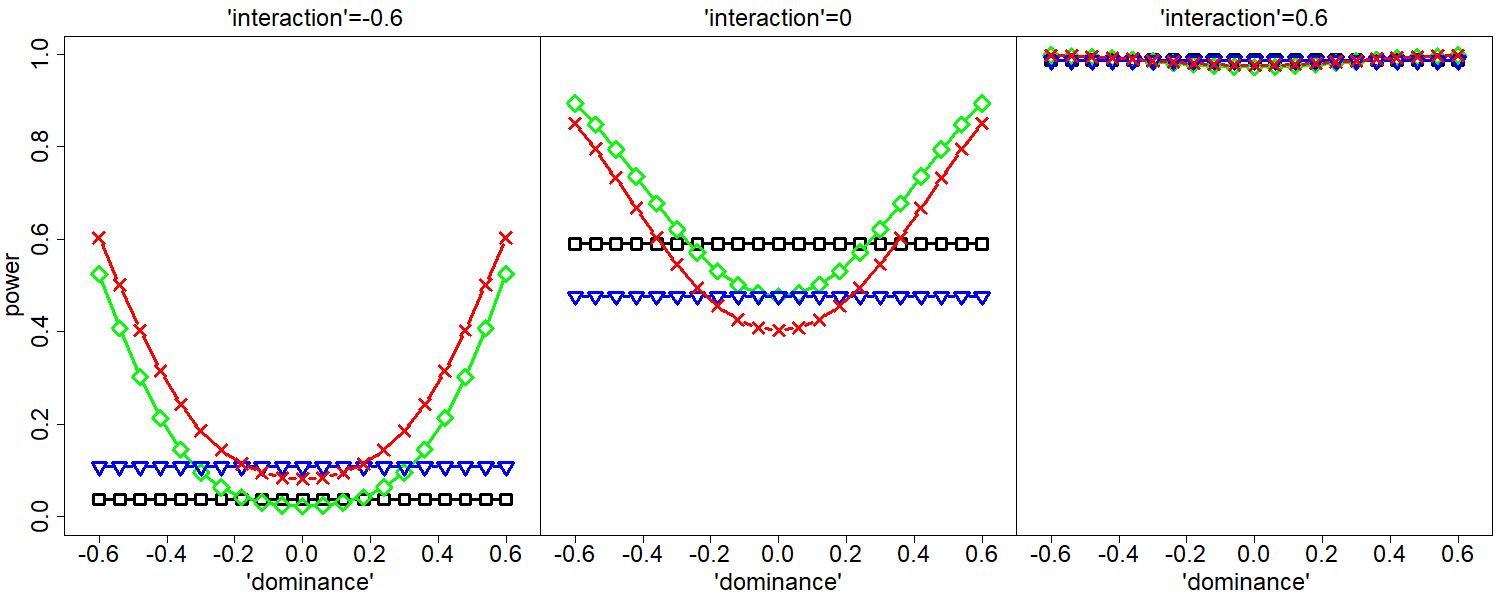}\end{center}
\begin{center}\includegraphics[scale=0.3]{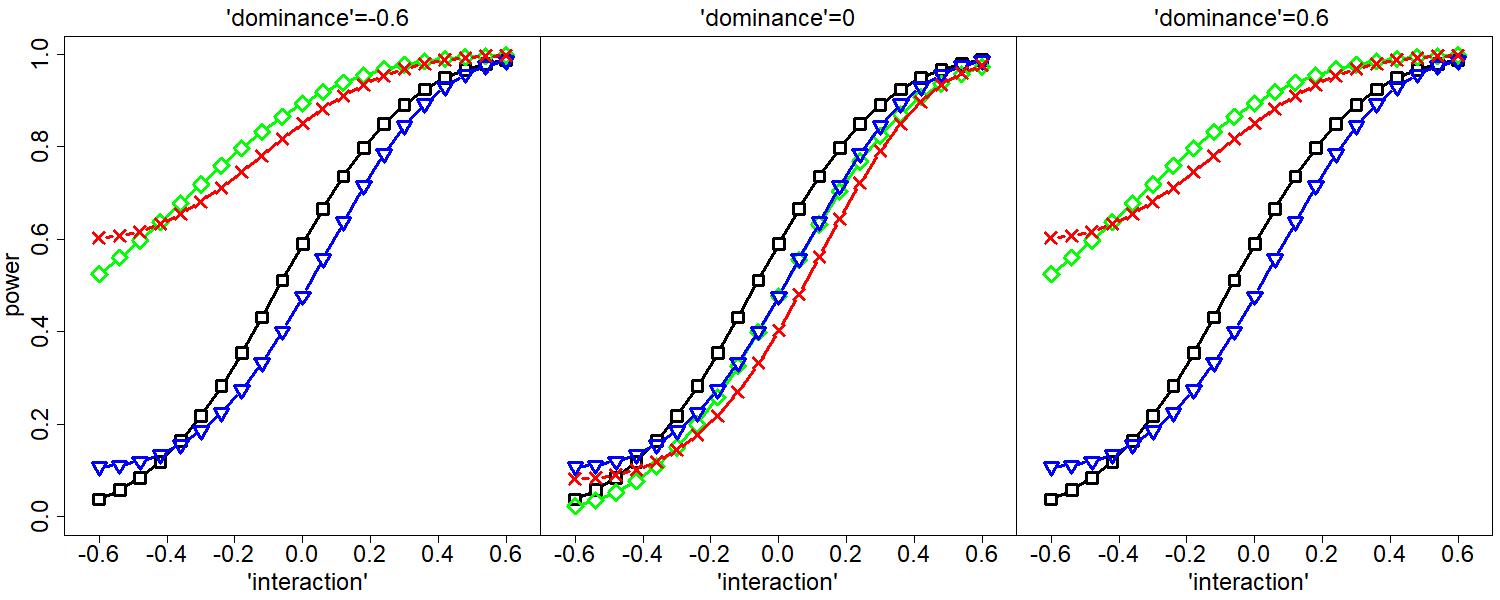}\end{center}
\end{figure}

\begin{figure}
\centering \textbf{E.} $\mathbf{f_{female}=0.8, f_{male}=0.2}$
\begin{center}\includegraphics[scale=0.3]{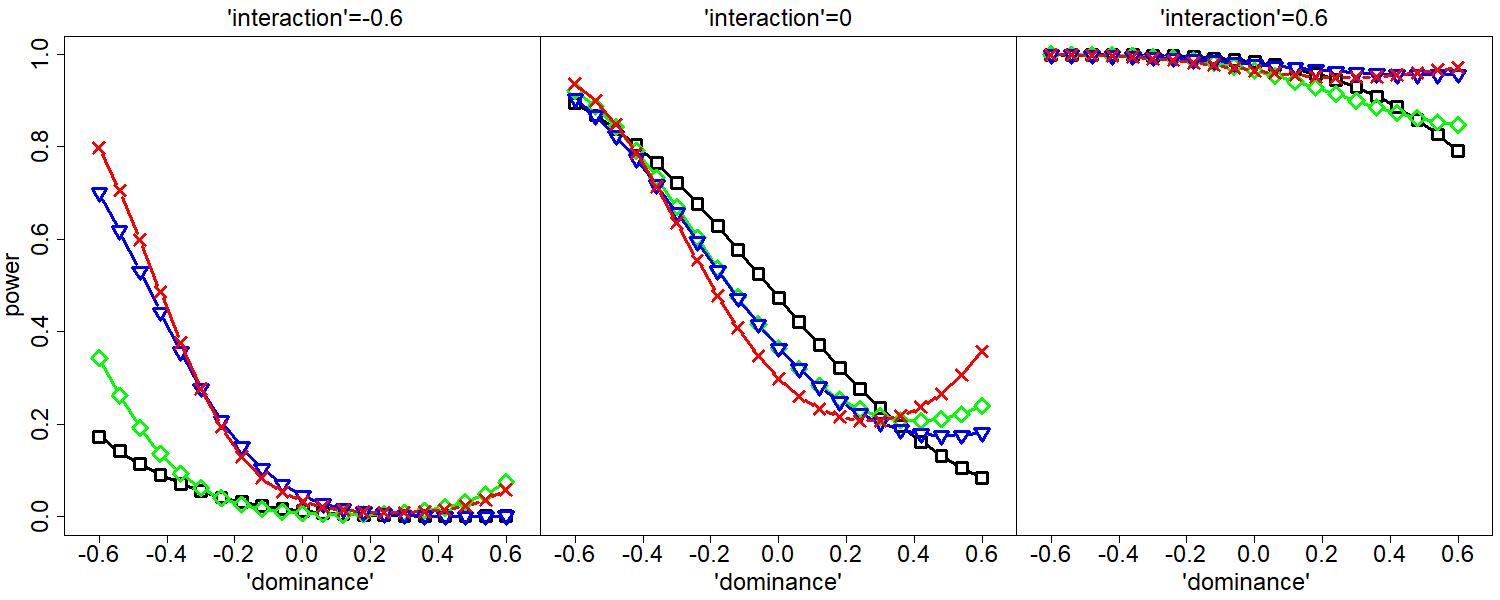}\end{center}
\begin{center}\includegraphics[scale=0.3]{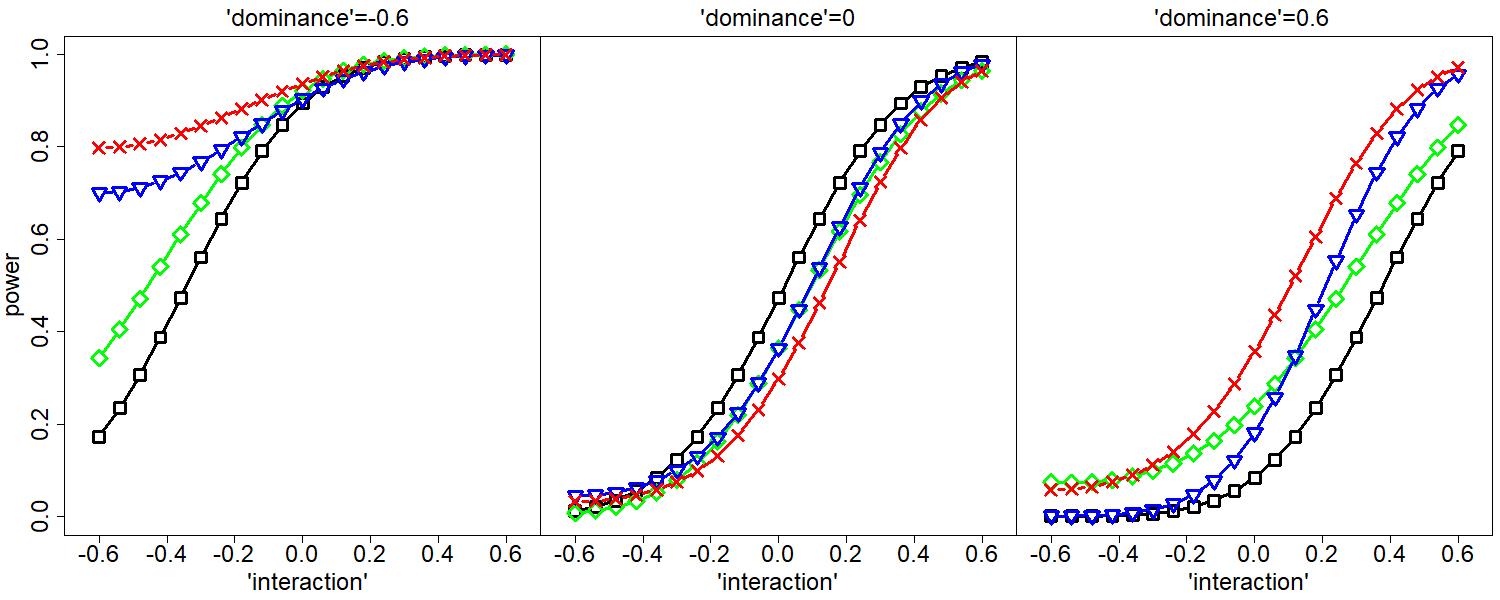}\end{center}
\end{figure}

\begin{figure}
\centering \textbf{F.} $\mathbf{f_{female}=0.8, f_{male}=0.5}$
\begin{center}\includegraphics[scale=0.3]{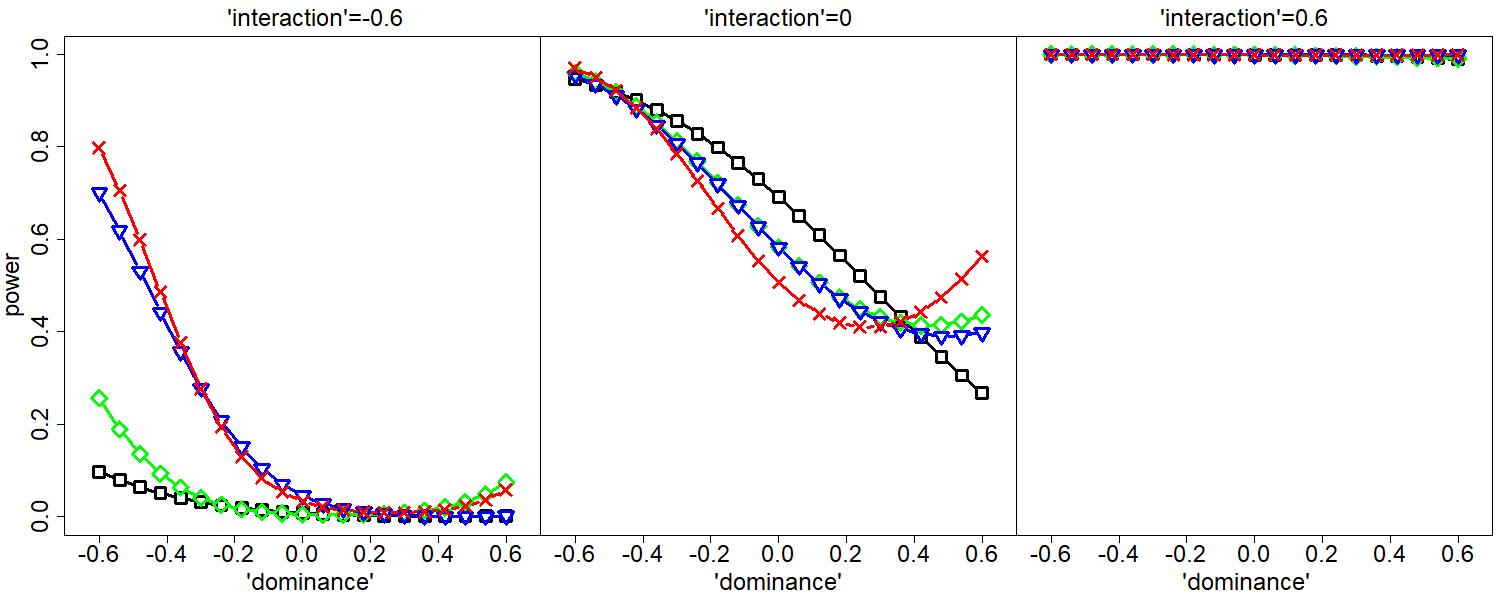}\end{center}
\begin{center}\includegraphics[scale=0.3]{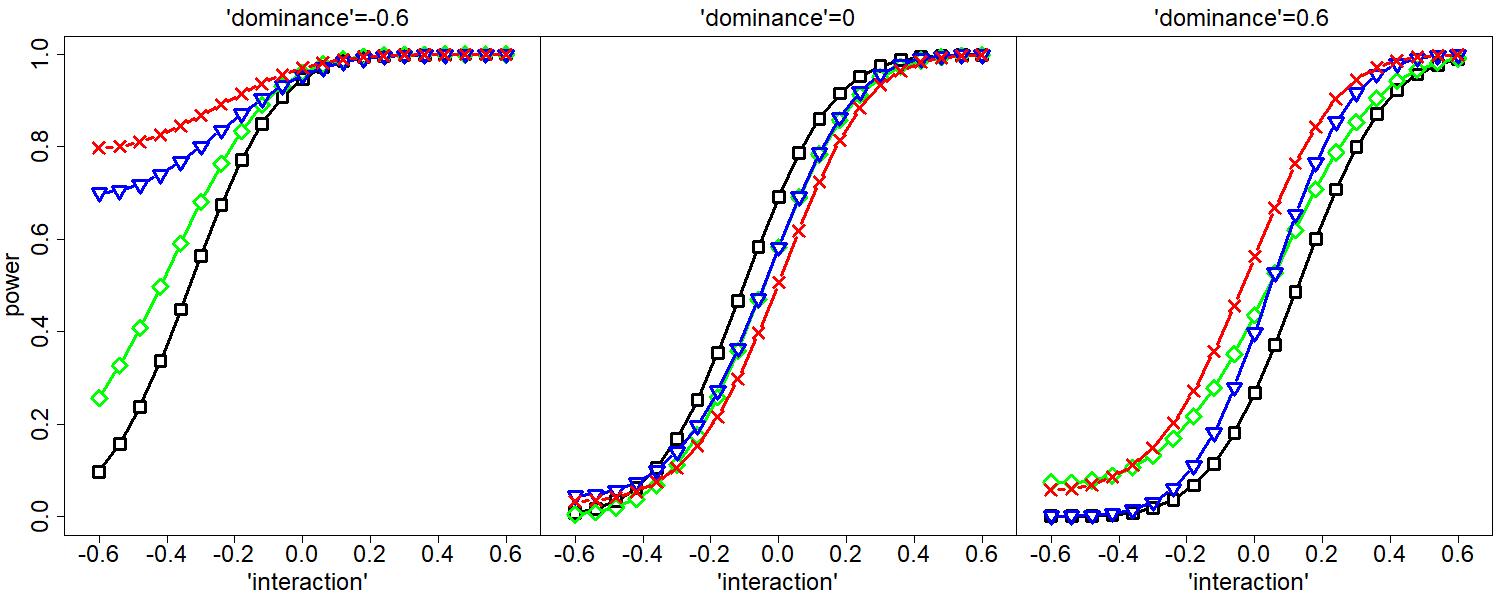}\end{center}
\end{figure}

\begin{figure}
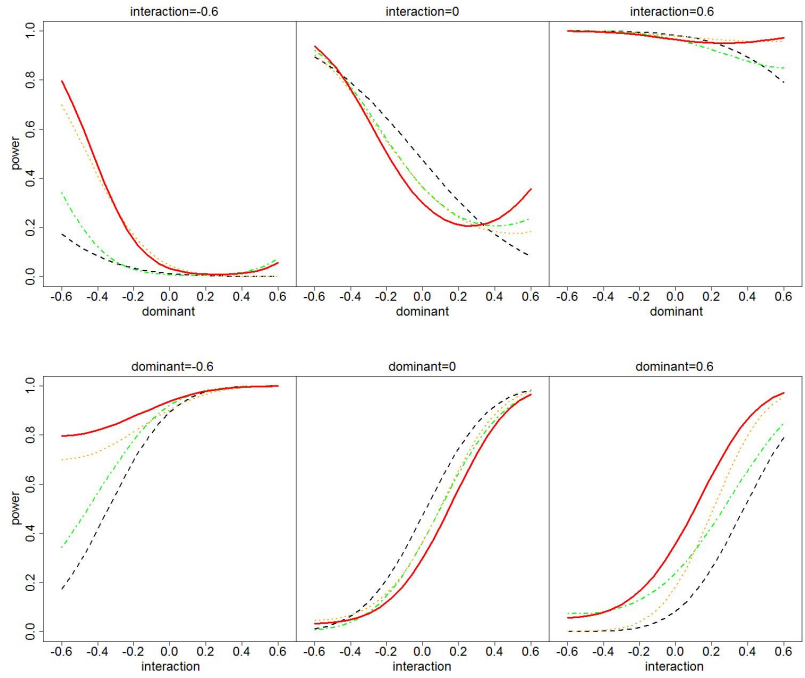

\centering \textbf{G.} $\mathbf{f_{female}=0.8, f_{male}=0.8}$
\begin{center}\includegraphics[scale=0.3]{88-1.jpeg}\end{center}
\begin{center}\includegraphics[scale=0.3]{88-2.jpeg}\end{center}
\caption{{\bf Power comparisons for analyzing X-chromosome SNPs.} Additional values of $f$ which are not presented in Figure 2 are specified through part A to G. {\color{black} Black $\square$ curves} for testing $\beta_A=0$ based on model $M_1$, {\color{green} green $\diamond$ curves} for testing $\beta_A=\beta_D=0$ based on model $M_2$, {\color{blue} blue $\triangledown$ curves} for testing $\beta_A=\beta_{GS}=0$ based on model $M_3$, and {\color{red} red $\times$ curves} for testing $\beta_A=\beta_D=\beta_{GS}=0$ based on the proposed model $M_4$.  \textbf{Upper panels} examine power as a function of dominant effect (or skewness of XCI). \textbf{Lower panels} examine power as a function of gene-sex interaction effect (or XCI status).}
\label{more powerX}
\end{figure}

\begin{figure}
\begin{center}\includegraphics[scale=0.6]{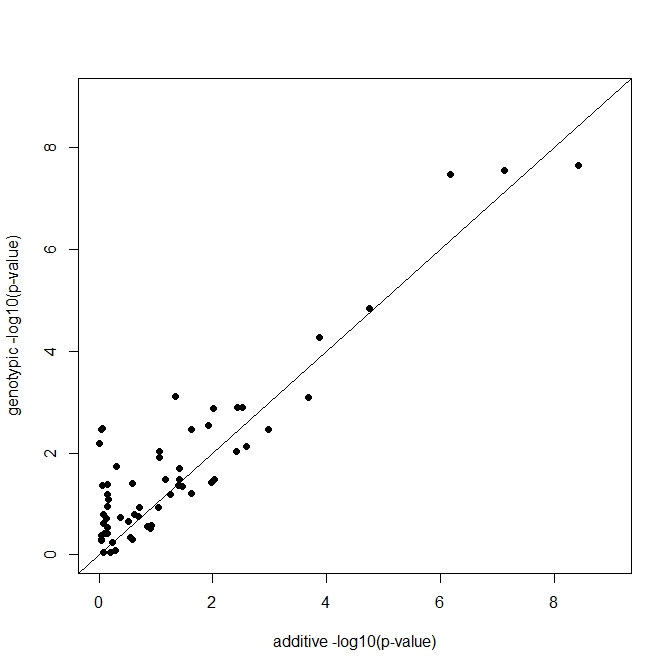}\end{center}
\caption{{\bf Results of re-analyses of the 60 autosomal, presumably associated, SNPs selected by \citet{wt05} from 41 association studies.} X-axis is the p-value, on the $-log_{10}$ scale, obtained from the standard 1 d.f.\ additive test and the Y-axis is the recommended 2 d.f.\ genotypic test.}
\label{pp60}  
\end{figure}

\begin{figure}
\begin{center}\includegraphics[scale=0.36]{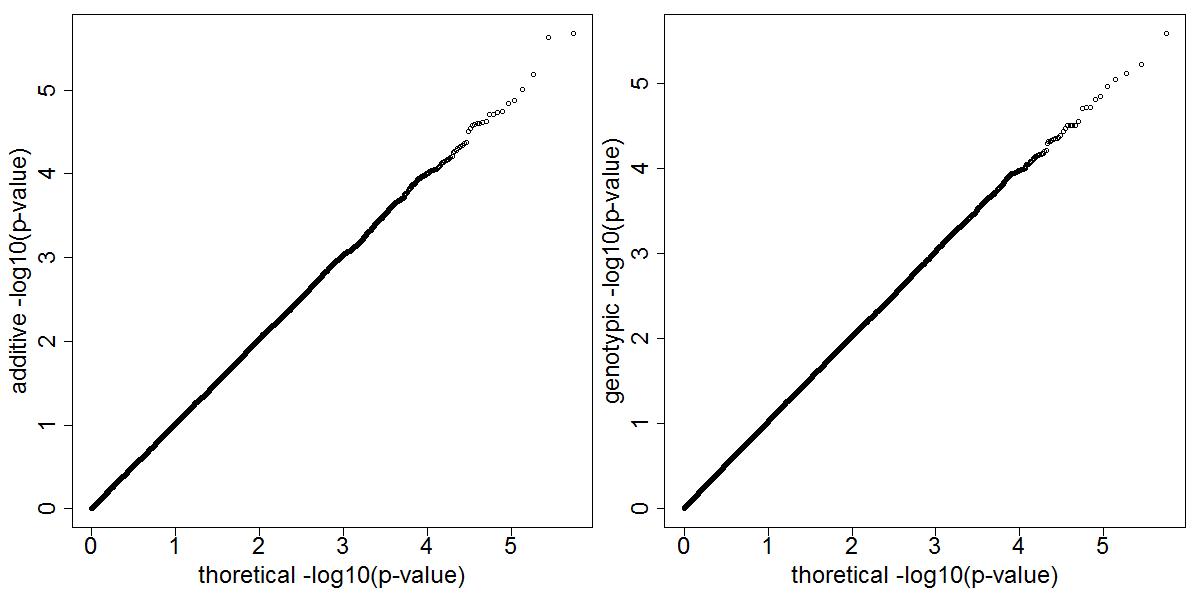}\end{center}
\caption{{\bf QQ-plots of the 556,445 autosomal SNPs from cystic fibrosis study in Section 4.1.} {\bf Left panel:} p-values of the additive test on $-log_{10}$ scale. {\bf Right panel:} p-values of the genotypic test on $-log_{10}$ scale. The QQ-plots imply that p-values are approximately Uniform(0,1) distributed for either test.}
\label{qq}  
\end{figure}

\begin{figure}
\begin{center}\includegraphics[scale=0.6]{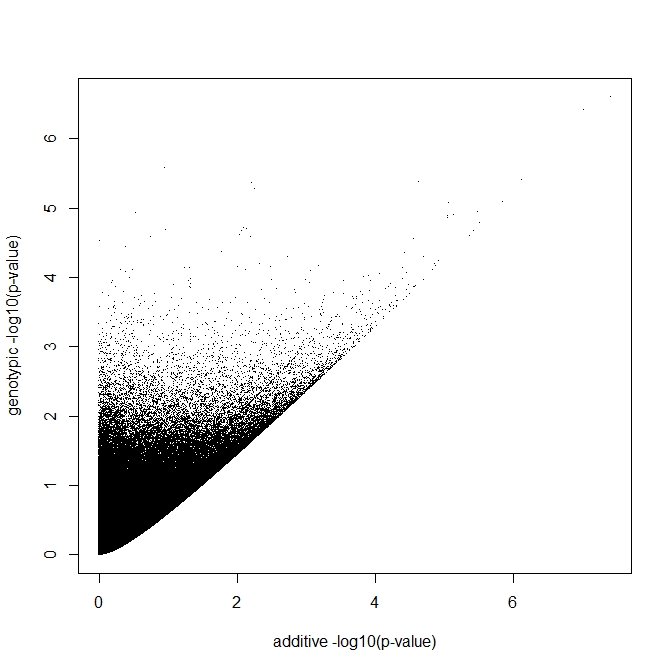}\end{center}
\caption{{\bf p-values of the additive test vs. genotypic test on $\mathbf{-log_{10}}$ scale of the 556,445 autosomal SNPs from cystic fibrosis study in Section 4.1.} Due to the capped maximal power loss computed in Section 3.1 and Appendix B, p-values under genotypic model is possible to be much smaller than additive model, but not possible to be much greater than additive model, which clearly demonstrates the benefit of genotypic model. It needs to be noted that the capped power loss does not contradict to the fact that the overall p-values under both additive and genotypic model have the same Uniform(0,1) distribution, as shown in Figure \ref{qq}.}
\label{pp}  
\end{figure}

\clearpage

\bibliographystyle{authordate1}

\bibliography{mybib}